\documentstyle[epsfig]{article} 
\newcommand{\figeps} [2]   {\centerline {\epsfig{file={#1}, height={#2}, clip=}}}  
\newcommand{\bea}{\begin{eqnarray}}
\newcommand{\eea}{\end{eqnarray}}
\def\beq{\begin{equation}}
\def\be{\begin{equation}}
\def\ee{\end{equation}}
\def\bd{\begin{displaymath}}
\def\ed{\end{displaymath}}
\def\defeq{\;\buildrel\hbox{\small def}\over{\,=}\;}    

\begin{document}
\vspace*{1.5cm}

\begin{center}
{\huge REGULAR MOTIONS IN EXTRA-SOLAR}\\
\bigskip
{\huge PLANETARY SYSTEMS}\\
\vspace*{2cm}
\bigskip
{\LARGE S.FERRAZ-MELLO,\,\,T.A.MICHTCHENKO}\\
\medskip
{\Large Instituto de Astronomia, Geof\'{\i}sica e Ci\^encias Atmosf\'ericas}\\
\smallskip
{\Large Universidade de S\~ao Paulo, Brasil}\\
\smallskip
{\large\rm (sylvio@usp.br, tatiana@astro.iag.usp.br)}\\
\bigskip
{\Large and}\\
\bigskip
{\LARGE{C.BEAUG\'E}}\\
\medskip
{\Large Observat\'orio Astron\'omico}\\
\smallskip
{\Large Universidad Nacional de C\'ordoba, Argentina}\\
\smallskip
{\large\rm (beauge@oac.unicor.edu)}\\
\end{center}
\bigskip
\bigskip

\begin{itemize}
\item[$\ $]
{\it Abstract.} This paper is a review of the dynamics of a system of planets. It includes the study of averaged equations in both non-resonant and resonant systems and shows the great deal of situations in which the angle between the two semi-major axes oscillates around a constant value. It introduces the Hamiltonian equations of the $N$-planet problem and Poincar\'e's reduction of them to $3N$ degrees of freedom with a detailed discussion of the non-osculating ``canonical'' heliocentric Keplerian elements that should be used with Poincar\'e relative canonical variables. It also includes Beaug\'e's approximation to expand the disturbing function in the exoplanetary case where masses and eccentricities are large. The paper is concluded with a discussion of systems captured into resonance and their evolution to symmetric and asymmetric stationary solutions with apsidal corotation.
\end{itemize}

\begin{itemize}
\item[$\ $]
\end{itemize}

\begin{itemize}
\item[$\ $]
To be published in "Chaotic Worlds: From Order to Disorder in Gravitational N-Body Systems" (B.A,Steves, ed.), Kluwer Acad. Publ. 
\end{itemize}

\vfill\eject

\section{Introduction}
The discovery of the first extra-solar planet in orbit around a main-sequence star was announced in 1995. 
Since then, the number of known extra-solar planets did not cease to grow. 
As the observations are accumulating, planetary systems with 2 and 3 planets are being discovered. 
More than ten are, presently, known\footnote{For an up-to-date list see the web page ``Extra-Solar Planets Encyclopaedia'', by J.Schneider at www.obspm.fr/planets and links therein.}. 
As the discoveries are recent and many of the discovered planets are at the edge of observational capabilities, the uncertainties on their orbital elements and masses are large. 
It is worth recalling that one of the 2-planet systems previously announced, HD 83443, vanished from the lists after new observations failed to show the radial velocity variations previously identified with a second planet. Another important example of the current uncertainties is the ``jump'' suffered by the determined eccentricity of HD 82943b. 
During long time, it was listed as $\sim 0.4$, while a new determination using observations over a long span of time gives only $0.18$. 
By the same occasion, the mass of HD 82943c became twice bigger than believed before (Mayor et al., 2004). 
These discrepancies should be enough to show us how hazardous is the task of getting conclusions from the present data and that we should avoid conclusions critically depending on the available data. 

In the current state of art, we are just capable of discovering big planets with not too large periods. Therefore, the planets so far discovered are big and most of them have orbits close to the central stars. 
Another characteristic is the large eccentricities of many of them. 
Even if large eccentricities favors discovery, this characteristic is not only due to observational bias and needs an explanation.
(See Perryman 2000 for a review of the existing hypotheses.) 
Large eccentricities are considered as the result of early migration processes.
It is generally believed that the planets did not form at their present observed locations, but were driven by a migration process due to tidal interaction of the planets with the discs where they were formed (see Papaloizou, 2003). 
Whether this orbital drift is still at work or not is a matter of debate, although it is more plausible to assume that it stopped after the end of the planetary formation stage. 
These early processes were also responsible for having driven the (surviving) systems to very stable conditions in which orbit periapses appear close to alignment or anti-alignment. 
This condition is observed in several systems.

Periapses alignment (or anti-alignment) may occur in resonant and non-resonant systems alike. In resonant systems, they are the natural states after the system is trapped into a mean-motion resonance (see section \ref{capture}). At variance, in non-resonant system, they are a consequence of the angular momentum  variations during resonance crossings without capture (Ferraz-Mello et al. in preparation).
However, and independently on how they reached this condition, an important consequence of this type of configuration is that they constitute a stabilizing mechanism for planetary orbits, especially if they have large eccentricities.

Four extra-solar systems seem to satisfy the resonance condition:
{Gliese 876}, { HD 82943}, {55 Cnc} and { 47 UMa}. The first two have planets with periods in a 2/1 commensurability, the third in a 3/1, and the later close to a 7/3. 
With regard to { Gliese 876}, numerical simulations (Laughlin and Chambers 2001, Lee and Peale 2002) seem to indicate that these bodies are actually deeply trapped in an apsidal corotation (see section \ref{apsidal}): They exhibit a libration of both resonant angles $\sigma_i = 2\lambda_2 - \lambda_1 - \varpi_i$, and also an alignment of their major axes. 
Apsidal corotation seems to be the natural issue of a capture in resonance in the case of two planets with initially low eccentricities (Ferraz-Mello et al. 2003; {\it cf.} this paper, section \ref{capture}). 
The alignment (or anti-alignment) of periapses has not yet been confirmed in the case of the other planetary systems above mentioned.

The most conspicuous non-resonant system showing nearly aligned periapses is $\upsilon$ Andromedae. This system has been the object of many numerical and analytical studies (for references, see Michtchenko and Malhotra, 2004). 
The orbit of the planets $c$ and $d$ in this system are such that the distance between their periapses oscillates about zero with half-amplitude $\sim 60$ degrees and period $\sim 7260$ years. 

\section{Hamiltonian Equations of the N-Planet Problem}

This section considers the Hamiltonian formulation of the problem of N planets orbiting a star in an arbitrary configuration.
This is a well-known problem in Celestial Mechanics. 
However, the vast majority of papers in Celestial Mechanics deal with the so-called restricted 3-body problems in which only 2 bodies have finite masses.
Therefore, some basic topics of the general problem need to be remembered.

\subsection*{Barycentric Hamiltonian Equations}

The barycentric Hamiltonian equations of the N+1 body problem are obtained using the basic principles of Mechanics. Let $m_i$ ($i=0,1,\cdots,N$) be their masses.
If we denote as ${\mathbf X}_i$ the position vectors of the N+1 bodies with respect to an inertial system, and ${\mathbf \Pi}_i=m_i\dot{\mathbf X}_i$ their linear momenta, these variables are canonical and the Hamiltonian of the system is nothing but the sum of their kinetic and potential energies:
\be
\tilde{H}=T+U=\frac{1}{2}\sum_{i=0}^{N} \frac{{\mathbf \Pi}_i^2}{m_i}
- G \sum_{i=0}^{N} \sum_{j=i+1}^{N} \frac{m_i m_j}{\Delta_{ij}}
\ee
where G is the constant of gravitation and $\Delta_{ij}=|{\mathbf X}_i - {\mathbf X}_j|$.
This system has, however, $3(N+1)$ degrees of freedom, that is, 6 equations more than the usual Laplace-Lagrange formulation of the heliocentric equations of motion.
The system can be reduced to 3N degrees of freedom through the convenient use of the trivial conservation laws concerning the inertial motion of the barycenter.
There are two sets of variables used to reduce to $3N$ the number of degrees of freedom of the above system. 
The most popular reduction, due to Jacobi, is widely used in the study of the general three-body problem and of planetary and stellar systems.
A less popular reduction is due to Poincar\'e;
it was first published in 1897, but Poincar\'e himself did not use it because of difficulties related with the definition of the associated Keplerian elements (see Poincar\'e, 1905; see next section).
It appeared in the literature from times to times and started being more frequently used around the eighties (Yuasa and Hori, 1979; Hori, 1985; Laskar, 1990).
Hagihara (1970) says that it was discovered by Cauchy.

\begin{figure}
\figeps{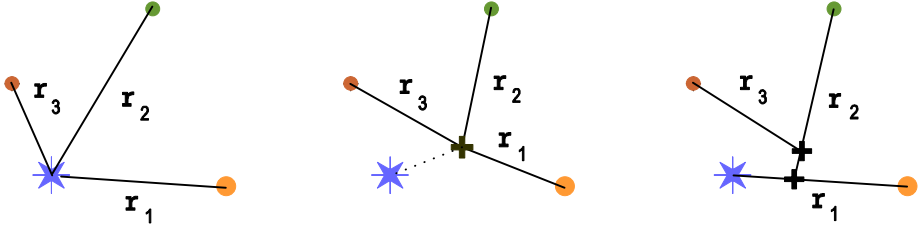}{3cm}
\caption{Main systems of coordinates: heliocentric (left), barycentric (center) and Jacobi's (right).}
\label{sl2-05}
\end{figure}

\subsection{Poincar\'e's Reduction to $3N$ degrees of freedom}

In Poincar\'e's reduction, the variables are the components of the heliocentric position vectors ${\mathbf X}_i - {\mathbf X}_0$ and the momenta are the same linear momenta ${\mathbf \Pi}_i$ of the barycentric formulation.
Hence, 
\be
{\mathbf r}_i = {\mathbf X}_i - {\mathbf X}_0,\hspace{2cm} {\mathbf p}_i = {\mathbf \Pi}_i,
\ee
$(i=1,2,\cdots,N)$. The given system has N+1 bodies and we thus need to introduce one more pair of (vector) variables.
Let them be
\be
{\mathbf r}_0 = {\mathbf X}_0,\hspace{2cm} {\mathbf p}_0 = \sum_{i=0}^{N}{\mathbf \Pi}_i.
\ee
A trivial calculation shows that the variables ${\mathbf r}_i, {\mathbf p}_i$ ($i=0,1,\cdots,N)$ are canonical.
Let us, now, write the Hamiltonian in terms of the new variables.
The transformations of $T$ and $U$ give, respectively,
\be
T=\frac{1}{2}\sum_{i=1}^{N} \frac{{p}_i^2}{m_i} +
\frac{1}{2}\sum_{i=1}^{N} \frac{{p}_i^2}{m_0} + 
\frac{1}{2} \frac{{p}_0^2}{m_0} 
- \sum_{i=1}^{N} \frac{{\mathbf p}_0 \cdot {\mathbf p}_i}{m_0} +
\sum_{i=1}^{N} \sum_{j=i+1}^{N} \frac{{\mathbf p}_i \cdot {\mathbf p}_j}{m_0}
\ee
and
\be
U= - G \sum_{i=1}^{N} \frac{m_0 m_i}{r_i}
- G \sum_{i=1}^{N} \sum_{j=i+1}^{N} \frac{m_i m_j}{\Delta_{ij}}
\ee
where $p_i = |{\mathbf p}_i|\,$ and $r_i = |{\mathbf r}_i| = |\Delta_{0i}|$.

The reduction of the system is immediate.
We note, beforehand, that the variable ${\mathbf r}_0$ is ignorable.
Consequently, ${\mathbf p}_0$ is a constant that, by construction, we set equal to zero.
The resulting equations may be separated into two parts:
\begin{itemize}
\item[A.] The first pair of equations, corresponding to the subscript $0$, is:
\be
\dot{\mathbf p}_0 = 0 \hspace{2cm}
\dot{\mathbf r}_0 = {\rm grad}_{{\mathbf p}_0} \tilde{H}.
\label{eq6}\ee
We note that the second of eqns.\,(\ref{eq6}) gives
\be
\dot{\mathbf r}_0 = \frac{{\mathbf p}_0}{m_0}
- \sum_{i=1}^{N} \frac{{\mathbf p}_i}{m_0}. 
\ee

\item[B.] The canonical equations in the variables ${\mathbf r}_i,\, {\mathbf p}_i , \, (i\neq 0)$ are given by the reduced Hamiltonian
\be
{H} = \tilde{H} - \frac{1}{2} \frac{{p}_0^2}{m_0} 
+ \sum_{i=1}^{N} \frac{{\mathbf p}_0 \cdot {\mathbf p}_i}{m_0}. 
\ee
This subsystem has 3N degrees of freedom and is separated from the previous one, since ${\mathbf p}_0$ is constant. (We did assume  ${\mathbf p}_0 = 0$.)
\end{itemize}
The Hamiltonian of the reduced system is 
${H} = {H}_0 + {H}_1 $
where
\be
{H}_0 =\sum_{i=1}^{N}\left( \frac{1}{2} \frac{{p}_i^2}{\beta_i} - \frac{\mu_i\beta_i}{r_i}\right)
\ee
\be
{H}_1 = 
\sum_{i=1}^{i=N} \sum_{j=i+1}^{j=N} \left( -\frac{Gm_i m_j}{\Delta_{ij}}
+ \frac{{\mathbf p}_i \cdot {\mathbf p}_j}{m_0} \right)
\label{Hpoi}\ee
and
\be
\mu_i = G (m_0+m_i) \hspace{2cm}
\beta_i=\frac{m_0m_i}{m_0+m_i}.
\label{mubetaPoi}\ee
We note that ${H}_0$ is of the order of the planetary masses $m_i$ while ${H}_1$ is of order two with respect to these masses.
Then ${H}_0$ may be seen as the new expression for the undisturbed energy while ${H}_1$ is the potential energy of the interaction between the planets.

It is worth noting that each term
\be
F_i = \frac{1}{2} \frac{{p}_i^2}{\beta_i} -  \frac{\mu_i\beta_i}{r_i}
\label{Fk-P}\ee
is the Hamiltonian of a two-body problem in which the mass-point $m_i$ is moving around the mass point $m_0$.
In fact, from the Hamiltonian given by eqn.\,(\ref{Fk-P}), it is easy to obtain the second-order differential equation
\be
\ddot{\mathbf r}_i = - \mu_i  \frac {{\mathbf r}_i}{r_i^3} = -  G (m_0+m_i)  \frac {{\mathbf r}_i}{r_i^3}.
\ee

One of the canonical equations spanned by $F_i$ is

\be
\dot{\mathbf r}_i = \frac{{\mathbf p}_i}{ \beta_i}.
\label{rdotPoi}\ee
This equation apparently contradicts the statements done after which ${\mathbf r}_i$ is the heliocentric radius vector and ${\mathbf p}_i$ is the barycentric linear momentum. 
However, it only means that the variation of ${\mathbf r}_i$ in the reference Keplerian motion is not the actual relative velocity of the $i^{\rm th}$ body but ${\mathbf p}_i/\beta_i$.
This means that, at variance with other formulations, the Keplerian motions defined by eqns.\,(\ref{Fk-P}) are not tangent to the actual motions. To distinguish them from ``heliocentric osculating'', when necessary, we will use the word ``heliocentric canonical''. 

\subsection{Action-angle variables. Delaunay elements}\label{Delaunay}

The solution of $H_0$ is a set of $N$ Keplerian motions whose generic Hamiltonian is $F_i$.
The purpose of this and the forthcoming section is to obtain the Keplerian elements and the Delaunay variables corresponding to the relative coordinates introduced before, which must be used when a canonical perturbations theory is constructed using $H_0$ as ``unperturbed'' approximation. 
For that sake, we have to solve the corresponding Hamilton-Jacobi equation and construct the action-angle variables of the given problem. 
We only give here the more important steps characterizing the variables appearing in the definitions of their action-angle variables and in the associated Delaunay elements. 
To do it, the study of the planar case is enough and preferable since the rotations necessary when the spatial case is considered, although trivial, introduce many new equations.
All conceptual questions appear in the planar case and have the advantage of making the calculations much easier and thus allow the crucial points to be clearly identified.
Once the conceptual problems are solved in the planar case, the usual three-dimensional equations can be easily adapted to give the elements we are looking for.
In the plane, the Hamiltonian is separable in polar coordinates. 
To introduce these variables, let us remember that, in the reference Keplerian motion,  ${\mathbf p} = \beta \dot{\mathbf r}$. (For the sake of simplicity, we omit the subscript $i$ in the forthcoming equations.)
Then 
\be
{\mathbf p} = \beta \left(
\dot{r} {\mathbf a} + r\dot{\psi} {\mathbf b}\right)
\ee
where ${\mathbf a}, {\mathbf b}$ are the right-handed set of unit vectors at $\mathbf r$ in the positive directions of the increments of $r,\psi$. $\,\dot{r},\dot\psi$ are the time derivatives of $r,\psi$ in the reference Keplerian motion.
The kinetic energy term is, then,
\be
{T}=\frac{\beta}{2}(\dot{r}^2 + r^2 \dot{\psi}^2 )
\ee
or, introducing the momenta $p_r = \frac{\partial T}{\partial \dot{r}}$ and $p_\psi = \frac{\partial T}{\partial \dot{\psi}}$, we obtain
\be
T = \frac{1}{2\beta}\left(p_r^2 + \frac{p_\psi^2}{r^2} \right).
\ee
The potential energy term is given by
\be
U(r) = - \frac{\mu\beta}{r}
\ee
and the resulting Hamilton-Jacobi equation is the classical one of the two-body problem with $\beta$ instead of $m$ and $\mu$ instead of $G(M+m)$: 
\be
F = \frac{1}{2\beta}\left(p_r^2 + \frac{p_\psi^2}{r^2} \right) - \frac{\mu\beta}{r}.
\ee

The solution of this equation is well known and does not need to be reproduced here with all details. This equation is separable into:
\be
p_r = \sqrt{2\beta(E + \frac{\mu\beta}{r}) - \frac{C^2}{r^2}}
\ee
\be
p_\psi = C.
\ee
$C,E$ are integration constants ($E=F$ is the ``energy" and  $C = {\mathbf r} \times {\mathbf p}$ is the ``angular momentum"; the quotation marks are necessary because of the particular definitions of ${\mathbf r}$ and ${\mathbf p}$ in the considered formulation).

The actions associated with the given Hamiltonian are
\be
J_r=\frac{1}{2\pi}\oint p_r dr
\hspace{2cm}
J_\psi=\frac{1}{2\pi}\oint p_\psi d\psi
\ee
whose integrations give
\be
J_r = -C + \mu\beta\sqrt{\frac{\beta}{-2E}} \hspace{2cm} J_\psi=C.
\label{Jr}\ee

The Delaunay actions are:
\be\begin{array}{l@{\ =\ }l@{\ =\ }l}
L & J_r + J_\psi & \beta \sqrt{\mu a} \\
G & J_\psi & \beta \sqrt{\mu a} \sqrt{1-e^2} \\
\end{array}\ee
where $a$ and $e$ are two constants introduced in the integration giving the action $J_r$:
\begin{itemize}
\item[$\bullet$] The mean distance (or semi-major axis) 
\be
a \defeq -\frac{\mu\beta}{2E}
\label{aKep}\ee
\item[$\bullet$] The eccentricity 
\be
e \defeq \sqrt{1 + \frac{2E C^2}{\mu^2\beta^3}}.
\label{eKep}\ee
\end{itemize}

Since, in general, the planets do not move in the same plane, we have to introduce the inclinations $I$ of their planes of motion over a fixed reference plane and add the third Delaunay action $H = \beta \sqrt{\mu a} \sqrt{1-e^2} \cos I $.
The Delaunay angles $\ell, \omega=\psi-v$ (and $\Omega$) are obtained in the usual way.
 
\subsection{Canonical heliocentric elements}

For each planet, we may transform $\beta, \mu, {\mathbf r}, {\mathbf p}$ into the elements $a, e, \lambda, \varpi $ using the same transformations used to define the ordinary osculating heliocentric elements $a_{\rm osc}, e_{\rm osc}, \lambda_{\rm osc}, \omega_{\rm osc}$ of the two-body problem as functions of $m, G(M+m), {\mathbf r}, {m\dot{\mathbf r}}$. However, the equations giving the osculating heliocentric elements depend on $m$ only through $\mu$.
In order to use always the same routines, the above equations may be transformed.
We substitute, in eqns.\,(\ref{aKep}) and (\ref{eKep}), $E$ and $C$ by their definitions $E=F$ and $C={\mathbf r}\times {\mathbf p}$.
We obtain the well-known equations
\be
a=\frac{\mu r}{2\mu - rw^2}
\label{aKep-w}
\ee
\be
e=\sqrt{\left(1-\frac{r}{a}\right)^2 + \frac{({\mathbf r}.{\mathbf w})^2}{\mu a}}
\label{eKep-w}
\ee
where we have used the velocity in the reference Keplerian motion
\be
{\mathbf w} = \frac {\mathbf p}{\beta},
\ee
instead of the actual planetary velocity, and $w=|{\mathbf w}|$.

The Keplerian motion corresponding to $H_0$ in Poincar\'e's relative canonical coordinates may be obtained with the ordinary routines substituting the heliocentric velocities by
\be
{\mathbf w} = \frac {m}{\beta}{\mathbf V}
\ee
where ${\mathbf V}$ is the absolute (i.e. barycentric) velocity.

The angles are obtained with usual equations. In the planar problem, the true longitude ($\phi$) is given by the angle formed by the radius vector with the first axis of the reference system (to be obtained through $\arctan y/x$ where $x,y$ are the components of $\mathbf r$). In the spatial problem, some rotations are necessary beforehand. The anomalies may also be easily obtained, starting with the eccentric anomaly ($u$), which is given by
\be
u = \arctan \left(\sqrt{\frac{a}{\mu}} \frac{{\mathbf r}.{\mathbf w}}{a-r}\right).
\ee   
The true ($v$) and mean ($\ell$) anomalies are obtained by means of classical 2-body equations.
The other angles to determine are the longitude of periapsis ($\omega = \phi- v $) and the mean longitude ($\lambda = \ell + \omega$).

The elements of the reference Keplerian orbit at the time $t$ are $a, e, \omega, \lambda$. Since the parameter $\lambda$ is variable, it is convenient to substitute it by the so-called ``mean longitude at the epoch'' $(\lambda_0$), which is the value of $\lambda$ at a standard ``epoch" $t_0$: 
\be
\lambda = \lambda_0 + n (t-t_0)
\label{l-epoch}
\ee 
where $n = \sqrt{\frac{\mu}{a^3}}$ is the mean-motion in the reference orbit.

\subsection{The Conservation of the Angular Momentum}\label{MomAng}

If the only forces acting on the N+1 bodies are their point-mass gravitational attractions, the angular momentum is conserved:
\be
{\cal L} = \sum_{i=0}^N m_i {\mathbf X}_i \times \dot{\mathbf X}_i
\ee
Since $\sum_0^N m_i{\mathbf X}_i = \sum_0^N m_i\dot{\mathbf X}_i = 0$, 
the above equation gives
\be
{\cal L} = \sum_{i=1}^N {\mathbf r}_i \times {\mathbf p}_i,
\label{calL}\ee
that is
\be
{\cal L} = \sum_{i=1}^N \beta_i \sqrt{\mu_i a_i (1-e_i^2)}\cdot
{\mathbf k}_i
\label{calLkepl}\ee
where ${\mathbf k}_i$ are the unit vectors normal to the Keplerian planes. 
This is an exact conservation law. In this equation $a_i$ and $e_i$ are not the usual heliocentric osculating elements but the canonical heliocentric elements defined by equations\,(\ref{aKep-w}) -- (\ref{eKep-w}) where ${\mathbf w}_i$ are the absolute velocities corrected by the factors $m_i/\beta_i$.

The conservation law given by eqn.\,(\ref{calL}) is also true if Jacobian coordinates are used.
However, the expression
\be
\widehat{\cal L} = \sum_{i=1}^N m_i \sqrt{\mu_i a_i (1-e_i^2)}.{\mathbf k}_i
\label{calLwide}\ee
where $a_i$ and $e_i$ are the heliocentric osculating elements
(Keplerian elements defined by eqns.\,(\ref{aKep-w}) -- (\ref{eKep-w}) with the heliocentric velocities ${\mathbf v}_i$ instead of ${\mathbf w}_i$) 
often found in the literature, is not a true conservation law. 
One may easily see that:
\be
\widehat{\cal L} = {\cal L} - \sum_{i=1}^N m_i {\mathbf X}_0 \times \dot{\mathbf X}_0
\ee
showing that the quantity $\widehat{\cal L}$ has in fact a variation of order ${\cal O}(m_i^2)$.

\subsection{Two-body Expansions}\label{twobody}

For the sake of future calculations, let us recall some series expansions of the two-body problem. These expansions are helpful in the task of writing computer codes for automatic expansion of $H_1$ and hold in all systems of elements founded on unperturbed Keplerian motions.

The first result to be recalled concerns the Fourier expansion of some functions of the radius vector and true anomaly. They are the convergent series 
\bea
\label{MNeq13}
\biggl({r \over a} \biggr)^n \cos{(kf)} &=& \sum_{j=0}^\infty 
                  \bigl( X_{j}^{n,k} + X_{-j}^{n,k} \bigr) \cos{(j\ell)} \\
\biggl({r \over a} \biggr)^n \sin{(kf)} &=& \sum_{j=0}^\infty 
                  \bigl( X_{j}^{n,k} - X_{-j}^{n,k} \bigr) \sin{(j\ell)} \nonumber
\eea
where the superscript $n$ may be either positive or negative. The coefficients $X_{j}^{n,k}$ are the Hansen coefficients (see Tisserand, 1960; Kaula, 1962). Hansen coefficients are functions of the eccentricity. They may be developed into power series of the eccentricities:  
\be
\label{MNeq14}
X_{j}^{n,k} = e^{\vert k-j \vert} \sum_{s=0}^\infty Y_{s+u_1,s+u_2}^{n,k} 
e^{2s}
\ee 
($u_1 = \max{(0,j-k)}$ and $u_2 = \max{(0,k-j)}$) 
where the numbers $Y_{s+u_1,s+u_2}^{n,k}$ are the Newcomb operators. Newcomb operators obey to some simple recurrence relations, which allow them to be easily calculated for all values of the indices (see Brouwer \& Clemence, 1961).

Introducing eqn.\,(\ref{MNeq14}) into eqn.\,(\ref{MNeq13}), we obtain, after some algebra,
\bea
\label{MNeq16}
\biggl({r \over a} \biggr)^n \cos{(kf)} &=& \sum_{i=0}^\infty
             \sum_{m=-\infty}^\infty B_{n,k,i,m} e^i \cos{(m\ell)} \\
\biggl({r \over a} \biggr)^n \sin{(kf)} &=& \sum_{i=0}^\infty 
             \sum_{m=-\infty}^\infty C_{n,k,i,m} e^i \sin{(m\ell)} \nonumber
\eea
where $B_{n,k,i,m}$ and $C_{n,k,i,m}$ are constant coefficients expressed as  functions of Newcomb operators. These coefficients, first calculated by Leverrier, do not depend on the orbital parameters and may be calculated once for all. They have some interesting properties. The most important of them is the d'Alembert property: $B_{n,k,i,m} = C_{n,k,i,m} = 0$ when $|m|<i$ or when $|m|-i$ is odd. 

The latest expansions are power series in $e$ and their convergence depend on the singularities of the analytic function $u=u(e,\ell)$, which are at $|e|=0.6627434\cdots$. This is the convergence radius of the given series (see Wintner, 1941).

\section{Expansion of the Disturbing Function}

The Hamiltonian equations in relative coordinates may be used to study the planetary motions. In analytical studies, once introduced the new variables, the next step is to write $H_1$ in terms of the Keplerian elements. A well-known approach to this problem is the classical Laplacian expansion of $H_1$ into a Fourier series in the angles and a power series in the eccentricities, which introduces the functions of the semi-major axes known as ``Laplace coefficients". Another expansion sometimes found in the literature uses the expansion of $\frac{1}{\Delta}$ in Legendre polynomials of the ratio of the distances of the two planets to the central star. These expansions work well in their domains of validity. The Laplacian expansion is a good approximation if the orbital eccentricities are small. However, the radius of convergence of the expansion decreases (see Ferraz-Mello, 1994) with the increase of the ratio $\alpha$ of the two semi-major axes. For $\alpha \sim 0.6$ the series is no longer convergent for eccentricities as small as $\sim 0.2$. The expansions with Legendre polynomials are more stringent: they may only be used in the study of well hierarchized systems where the ratio of the distances of the perturbed and perturbing bodies to the central body remain small forever. This is the case of the lunar theory, in which the motion of the Moon around the Earth is disturbed by the Sun. Otherwise, the convergence of the expansion in Legendre polynomials is very slow and its use in planetary problems accounts for many wrong results. We present, in this lecture, an improvement of the technique first developed by Beaug\'e (1996). This ``expansion" is valid in large domains of the phase space excluding a domain around the singularities associated to collisions between the two bodies. In Beaug\'e's approximation, the number of terms necessary to represent $H_1$ depends on the magnitude of $H_1$ in the domain to be studied: Near the minimum of $|H_1|$, a few terms are enough to have a good representation. This number increases quickly as we approach orbits that may come close to a collision. At variance with Beaug\'e's early expansion, the present one (Beaug\'e \& Michtchenko, 2003) has no explicit restrictions with regard to eccentricities and inclinations.  

\subsection{Beaug\'e's approximation. The parameter $\delta$}

The big problem in the expansion of $H_1$ comes from the terms having $\Delta$ in denominator. In heliocentric coordinates, we can write:
\be
\label{MNeq7}
{1 \over \Delta} = \bigl( r_1^2 + r_2^2 - 2 r_1 r_2 \cos{S} \bigr)^{-1/2}
\ee
where $S$ is the angle between both bodies as seen from the central mass.
Introducing the ratio $\rho = r_1/r_2$, eqn.\,(\ref{MNeq7}) becomes
\be
\label{MNeq8}
{r_2 \over \Delta} = \bigl( 1 + \rho^2 - 2 \rho \cos{S} \bigr)^{-1/2} .
\ee
Instead of expanding this function in Fourier series of $S$ (Laplace approach) or power series of $\rho$ (Legendre polynomials), a best-fit approach is used. We write
\be
\label{MNeq9}
{r_2 \over \Delta} = \bigl( 1 + x \bigr)^{-1/2}  .
\ee
where 
\be
x = \rho^2 - 2\rho\cos{S}
\ee
and represent the function $(1 + x)^{-1/2}$ by a polynomial of order $N$ in $x$: 
\be
\label{MNeq9bis}
(1 + x)^{-1/2} \simeq \sum_{n=0}^N b_n x^n
\ee
whose coefficients $b_n$ are determined numerically through a linear regression.

\begin{figure}[t]
\figeps{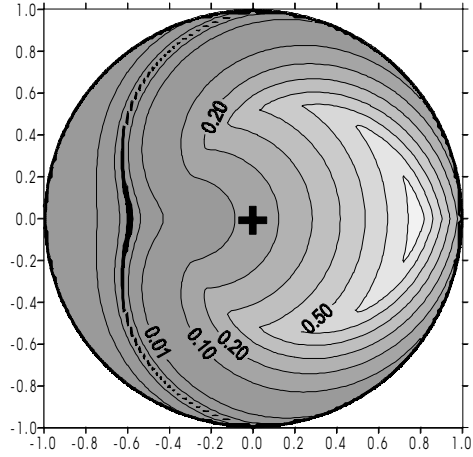}{6.3cm}
\caption{Limits of validity of Beaug\'e's approximations for planets in the 2:1 resonance with $e_2=0$ for different values of $\delta$. 
The thick black line on the left-hand side is the locus of the points where $x_{\rm min}=-1$ (collision curve). 
The non-labeled curves adjacent to it correspond to $\delta=0.001$. 
Horizontal axis: $e_1 \cos{\sigma_1};$ 
Vertical axis: $e_1 \sin{\sigma_1}.$ 
$\,(\sigma_1 = 2\lambda_2 - \lambda_1 - \varpi_1)$.} 
\label{MNfig2}
\end{figure}

The variable $x$ is a measure of the proximity of the initial condition to the singularity in $1\over\Delta$. It is equal to $-1$ at the singularity, and takes values larger than this for every point ($\rho, \cos{S}$) away from  the collision curve (see fig. \ref{MNfig2}). We note that the values of $\rho$ and $S$ are not separately significant; only the distance from the singularity is important.

The numerical fit is performed using values of $x > -1 + \delta$, where $\delta$ is a positive parameter close to zero. The smaller its value, the better the approximation to the real function near the singularity. However, when $\delta$ is small, the number $N$ of terms to be considered in the representation of $(1 + x)^{-1/2}$ to guarantee an adequate precision for all values of the independent variable is necessarily large. 

\begin{figure}
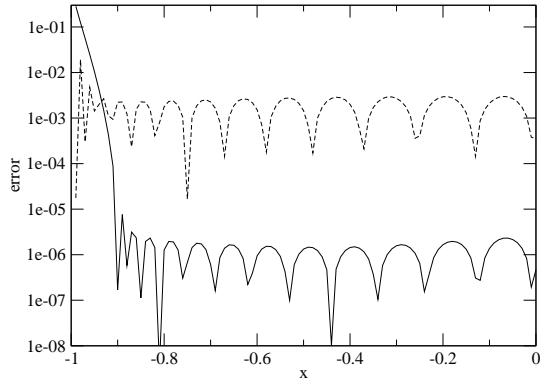

\figeps{rou1mn.eps}{5cm}
\caption{Relative error of the approximation of $1/\sqrt{1+x}$ given by eqn.\,(\ref{MNeq9bis}) 
in function of $x$, for two values 
of $\delta$. The continuous line shows the case $\delta=0.1$ and the broken line 
shows $\delta=0.01$. In both examples, $N=30$.}
\label{MNfig1}
\end{figure}

Figure \ref{MNfig1} shows the relative error of (\ref{MNeq9bis}) for $N=30$ and two values of $\delta$. We can see that for most of the interval of $x$, the fit with $\delta=0.1$ yields a much higher precision than the fit with $\delta=0.01$. In the fit with $\delta=0.1$, the errors are of the order of $10^{-6}$, that is, about 3 orders of magnitude lower than in the other case. Conversely, as $x \rightarrow -1$, the fit with $\delta=0.01$ is more precise. Larger values of $N$ will diminish the error in both cases, but at the cost of increasing the number of terms enormously. 

In the general case, the motion of the two bodies is unconstrained and the distance between the two planets is minimum in a symmetric conjunction with the outer planet at the periapsis and the inner planet at apoapsis. In this case, we have to choose $\delta<(1-\xi)^2$ where $\xi=\alpha(1+e_1)(1-e_2)^{-1}$. Beaug\'e's technique no longer requires that the ratio of the distances of the two planets is small, but it requires $\xi<1$. However, when the planets are in resonant motion, the method is valid even for crossing orbits because the resonance does not allow the planets to come close one to another. The limits of $x$ when the motion of the two planets is constrained by a 2:1 commensurability ($\alpha=0.63$) are shown in figure \ref{MNfig2} in the particular case where $e_2=0$.

The geometry of the curves in fig. \ref{MNfig2} follows very closely (but not identically) the topology of the phase portrait of the 2:1 resonant restricted three-body problem averaged over short-period terms. 
The maximum value of $x_{\rm min}$ lies at $e_1=0.8, \sigma_1=0$ (on the horizontal axis) and corresponds to the minimum of $|H_1|$. This point is very close to the corotation stationary solution of the 2:1 asteroidal resonance ($e_1=0.73$ when $e_2\rightarrow 0$; see Ferraz-Mello et al., 1993). Similarly, the minimum value of $x_{\rm min}$ (equal to $-1$) corresponds to the singularities of $H_1$. 
There is no direct relationship between the eccentricity and $x_{min}$. 
An orbit with a large eccentricity near the corotation center may have a larger value of $x_{min}$, while an almost circular orbit with a lower eccentricity may reach values very close to the limit $x=-1$. 
It is worth recalling that several extra-solar planet pairs observed in resonant configuration lie near corotation centers where $x_{min}$ is large and good Beaug\'e's approximations may be obtained with small $N$.

\subsection{The Direct Part}

To transform the above approximation into a function having the form needed in a theory, many transformations have to be done.
Introducing the explicit expression for $x$ into eqn.\,(\ref{MNeq9bis}), it becomes
\be
\label{MNeq10}
{r_2 \over \Delta} \simeq \sum_{k=0}^N \sum_{j=0}^n c_k (-2)^j 
\left( \begin{array}{c} k \\ j \end{array} \right) \rho^{2k-j} \cos^j{S}
\ee
where the $c_k$ are constant coefficients, easily obtainable in terms of the $b_k$. 

From now on, we will restrict ourselves to coplanar orbits.
Changing from powers of the cosines to multiples of the argument, and introducing the planar approximation $ S=f_1-f_2+\Delta\varpi$, we can rewrite it as:
\be
\label{MNeq11}
{a_2 \over \Delta} \simeq \sum_{k=0}^N \sum_{i=0}^{N-k} 2 A_{k,i} \alpha^m
\biggl( {r_1 \over a_1} \biggr)^m  \biggl( {r_2 \over a_2} \biggr)^{-m-1}  
\cos{k(f_1-f_2+\Delta\varpi)}
\ee
where $m=2i+k$. 
 
At last, introducing eqn.\,(\ref{MNeq16}) into the expression of the direct part of the disturbing function, and reordering the terms, we get:
\be
{a_2 \over \Delta} \simeq \sum_{j,k=0}^\infty \sum_{m,n=-\infty}^\infty
\sum_{l=0}^N \sum_{i=0}^{N-l} A_{l,i} D_{2i+l,j,k,m,n}
     \alpha^{2i+l} e_1^i e_2^j \cos{(m\ell_1 - n\ell_2 + l\Delta \varpi)}
\label{MNeq17}
\ee
where the coefficients $D_{2i+l,j,k,m,n}$ are given by:
\bea
\label{MNeq18}
D_{2i+l,j,k,m,n} = 
{1 \over 2 \gamma_m \gamma_n} &  (B_{2i+l,l,j,|m|} + {\rm sign}(m)C_{2i+l,l,j,|m|})\times \hspace{1cm}\\
&(B_{-2i-l-1,l,k,|n|} + {\rm sign}(n)C_{-2i-l-1,l,k,|n|}) \nonumber
\eea
and $\gamma_m$ is a simple bi-valuated function defined as:
\bea
\label{MNeq19}
\gamma_m = \left\{ \begin{array}{cl} 
       1/2 &  {\rm if} \hbox{\hspace*{0.5cm}} m=0   \\  
	1  &  {\rm if} \hbox{\hspace*{0.5cm}} m>0 . \\  
       \end{array} \right .
\eea

Eqn.\,(\ref{MNeq17}) multiplied by the factor $\frac {G m_1 m_2} {a_2}$ gives the term of the direct part corresponding to the given pair of planets.

\subsection{The Indirect Part}

In Poincar\'e heliocentric relative coordinates, the indirect part of $H_1$ is
(see eqn.\,(\ref {Hpoi})):
\be
T_1=\sum_{i=1}^N \sum_{j=i+1}^N \frac{{\mathbf p}_i{\mathbf p}_j}{m_0}.
\ee
The linear momenta ${\mathbf p}_i$ may be obtained from the derivatives of 
the vector radii ${\mathbf r}_i(t), {\mathbf r}_j(t)$ in the Keplerian reference orbit (see eqn.\,(\ref{rdotPoi})). Then, 
\be
T_1=\sum_{i=1}^N \sum_{j=i+1}^N \beta_i \beta_j \frac{\dot{\mathbf r}_i(t)\dot{\mathbf r}_j(t)}{m_0}
\ee
or
\be
\label{MNeq21}
T_1 = \frac{\beta_1 \beta_2}{m_0}n_1 n_2 
\biggl[
{\partial x_1 \over \partial \ell_1} 
{\partial x_2 \over \partial \ell_2}  + 
{\partial y_1 \over \partial \ell_1} 
{\partial y_2 \over \partial \ell_2} 
\biggr]
\ee
where $\ell_i$ are the mean anomalies and $n_i$ the mean motions.
$x_i$ and $y_i$ are the components of ${\mathbf r}_i$ and are given by $x_i=r_i\cos(f_i+\varpi_i)$ and $y_i=r_i\sin(f_i+\varpi_i)$. 

In the sequence, we substitute the mean motions by the values issued from Kepler's third law and put into evidence the same factor used at the end of the previous section. Hence
\be
\label{MNeq21bis}
T_1 = \frac{G m_1 m_2}{a_2} {\cal A}\alpha^{-1/2} 
\biggl[
{\partial \over \partial \ell_1} \biggl( {x_1 \over a_1} \biggr)
{\partial \over \partial \ell_2} \biggl( {x_2 \over a_2} \biggr) + 
{\partial \over \partial \ell_1} \biggl( {y_1 \over a_1} \biggr)
{\partial \over \partial \ell_2} \biggl( {y_2 \over a_2} \biggr) 
\biggr].
\ee
${\cal A} =
\sqrt{\frac{\beta_1 \beta_2}{m_1 m_2}} \approx 1-\frac{m_1+m_2}{2m_0}$ 
is taken hereafter equal to 1, introducing an error of third order in the planetary masses. 
Using the expansions given in section \ref{twobody}, there follows
\bea
\label{MNeq23}
{x_1 \over a_1} &=& \sum_{i=0}^{\infty} \sum_{j=-\infty}^{\infty} I_{i,j}
                     e^i \cos{(j\ell_1 + \varpi_1)} \\
{y_1 \over a_1} &=& \sum_{i=0}^{\infty} \sum_{j=-\infty}^{\infty} I_{i,j}
                     e^i \sin{(j\ell_1 + \varpi_1)} \nonumber
\eea
where
\be
\label{MNeq23bis}
I_{i,j} = {1 \over 2 \gamma_j} \biggl( B_{1,1,i,|j|} + 
                               {\rm sign}(j)C_{1,1,i,|j|} \biggr)  .
\ee
After the differentiation of these equations with respect to the mean anomalies, and substitution in $T_1$, we obtain
\be
\label{MNeq24}
T_1  = {G m_1 m_2 \over a_2 \alpha^{1/2} } \sum_{j,k=0}^{\infty}
      \sum_{m,n=-\infty}^{\infty} mn I_{j,m} I_{k,n}
 e_1^j e_2^k \cos{(m\ell_1 - n\ell_2 + \Delta \varpi)} . 
\ee
Notice that, except for the dependence on $\alpha$, this series is formally similar to that giving the direct part of $F_1$. To complete the similarity, we can substitute the factor $\alpha^{-1/2}$ by a power series expansion in the neighborhood of the exact resonant value and write it as
\be
\label{MNeq25}
\alpha^{-1/2} = \sum_{i=0}^{2N} \overline{A}_i \alpha^i
\ee
where $\overline{A}_i$ are constant coefficients. 
With this change, $T_1$ now reads:
\be
\label{MNeq26}
T_1 = {G m_1 m_2 \over a_2} \sum_{i=0}^{2N} \sum_{j,k=0}^{\infty}
      \sum_{m,n=-\infty}^{\infty} \overline{A}_i mn I_{j,m} I_{k,n} 
     \alpha^i e_1^j e_2^k \cos{(m\ell_1 - n\ell_2 + \Delta \varpi)},
\ee
which is the final expression for the indirect part of the disturbing potential.

\subsection{The Disturbing Function}
Since both parts are formally similar, we can unify both expressions and obtain a single series for the disturbing function of the planetary three-body problem in heliocentric relative coordinates:
\be
\label{MNeq27}
H_1 = \frac{G m_1 m_2}{a_2} \sum_{j,k=0}^{\infty} \sum_{m,n=-\infty}^{\infty}
          \sum_{l=0}^{N} \sum_{i=0}^{2N} R_{i,j,k,m,n,l} \alpha^i
     e_1^j e_2^k 
          \cos{(m\ell_1 - n\ell_2 + l\Delta \varpi)} 
\ee
where $\alpha={a_1 \over a_2}$ and 
\be
\label{MNeq28}
R_{i,j,k,m,n,l} = A_{l,(i-l)/2} D_{i,j,k,m,n} 
                - \delta_{l,0} \overline{A}_i mn I_{j,m} I_{k,n}  ;
\ee
$\delta_{l,0}$ is Kronecker's delta function. 
Note that these coefficients are constant for all initial conditions, and therefore need only be determined once. (For more details, see Beaug\'e \& Michtchenko, 2003.) 

It is important to note that each term in $H_1$ depends on the mean anomalies $\ell_i$ and on the difference of the periapses longitudes $\Delta\varpi$. This means that if the arguments are written in terms of longitudes  $\lambda_i, \varpi_i$ only, they become $\kappa_1\lambda_1+\kappa_2\lambda_2+\kappa_3\varpi_1+\kappa_4\varpi_2$ with $\sum\kappa_i=0$. 
That is, $H_1$ is invariant to rotations of the reference frame.

\section{Secular Dynamics of 2 Planets}

The study of the secular dynamics is the study of the secular part of the Hamiltonian, obtained after an averaging over the mean longitudes. We will restrict ourselves in this text to the case of only two planets. To the first-order of the masses, the averaged Hamiltonian is the mean value of H:
\be
<H>\; = \frac{1}{4\pi^2} \int_0^{2\pi} \int_0^{2\pi} H d\lambda_1 d\lambda_2
\ee
or
\bd
<H>\; = - \sum_{i=1}^2 \frac {\mu_i^2 \beta_i^3}{2L_i^2} - R_{\rm sec} (L_i, I_i, \Delta\varpi)
\ed
where we have introduced the Delaunay variables $L_i, I_i=L_i-G_i$ defined in section \ref{Delaunay}.
Because of the invariance of $H_1$ with respect to rotations, once the $\lambda_i$ are averaged out, only terms with arguments  $\kappa_3\varpi_1+\kappa_4\varpi_2$ with $\kappa_3=-\kappa_4$ can remain in $R_{\rm sec}$. That is, $<H>$ depends on only one angle, the difference $\Delta\varpi$.
This means that the averaged equations have three ignorable angles, that is, three first integrals (conservation laws). They are
\bd\begin{array}{l}
L_1={\rm const.}\\
L_2={\rm const.}\\
K_2=I_1+I_2={\rm const}.
\end{array}\ed

\begin{figure}[t]
\centerline{
\epsfig{file=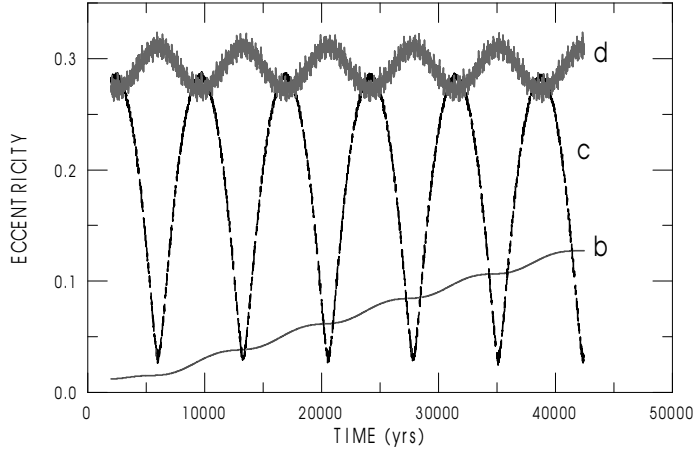,height=6cm,clip=}}
\caption{Secular variation of the osculating eccentricities of $\upsilon$And planets. The eccentricities of the two largest planets have variations in anti-phase as necessary to have $ {\rm AMD} \approx {\rm const.}$}
\label {ups}
\end{figure}

The third of these integrals, 
\be
K_2 = I_1 + I_2 = L_1 (1-\sqrt{1-e_1^2}) + L_2 (1-\sqrt{1-e_2^2}).
\label{amd}\ee
was called Angular Momentum Deficit by Laskar (2000). It is a combination of the conservation of the angular momentum ($G_1+G_2$=const., in the planar case) and the secular invariance of the $L_i$. The $I_i$ are positive quantities increasing from $0$ when $e_i=0$ to $L_i$ when $e_i=1$, that is, $0 < I_i < L_i$.
Therefore, in a system formed by only two planets, the $I_i$ shall vary in contrary directions and so shall vary the eccentricities: When one eccentricity increases, the other decreases (see fig. \ref{ups}).

This conservation law has some algebraic consequences. 
Assuming that $a_1 < a_2$, we have the following possibilities:
\begin{itemize}
\item[$\bullet$]{$\quad K_2 < L_1 < L_2 $}
\begin{enumerate}
\item $ I_1$ and $I_2$ cannot reach their maximum values $L_1$ and $L_2$, resp.;
\item $ e_1<1$ and $e_2<1$ (for all t);
\end{enumerate}
\item[$\bullet$]{$ \quad L_1 < L_2 < K_2$}
\begin{enumerate}
\item The AMD does not bound the eccentricities (both can reach $e=1$);
\item $e_1>0$ and $e_2>0$ (for all $t$).
\end{enumerate}
\item[$\bullet$]{$\quad L_1 < K_2 < L_2$}
\begin{enumerate}
\item $I_1$ can reach its maximum value $L_1$;
\item $I_2$ cannot reach its maximum value $L_2$ ($I_2 < L_2-L_1$);
\item The AMD does not bound $e_1$ (it can reach $e_1=1$);
\item The AMD bounds $e_2$ ($e_2<1$ for all $t$);
\item $I_2 > K_2 - L_1 > 0$ ;
\item $e_2>0$ (for all $t$).
\end{enumerate}
\end {itemize}

The conservation law is also found in N-planet systems. In the coplanar case, the angular momentum deficit is
\bd
K_N = I_1 + I_2 + \cdots + I_N = \sum_1^N L_i(1-\sqrt{1-e_i^2}).
\ed
It is worth emphasizing that this conservation law of the averaged system is not a rigorous one as the Angular Momentum conservation discussed in section \ref{MomAng}. It is approximated and valid strictly only as far as the hypotheses done to average the system are satisfied and the semi-major axes remain approximately constant. 

The equations of motion derived from $<H>$ are
\be
\dot{I}_1=-\frac{\partial <H>}{\partial \Delta\varpi}, \hspace{2cm} 
\Delta \dot{ \varpi}=\frac{\partial <H>}{\partial I_1}. 
\label{eqsec}
\ee
This system has only one degree of freedom and is integrable.

\subsection{The Mode I and Mode II Periodic Motions}
 
The exact solution of eqns.\,(\ref{eqsec}) is not easy to obtain analytically, but some insight can be gained by examining their equilibrium points (which correspond to periodic solutions of the two-degrees-of-freedom Hamiltonian $<H>$). They are defined by
\bea
\dot{I}_1=0{\rm ,}& &\Delta\dot{\varpi}=0.
\label{onedeg}\eea

For non-singular $I_1$ ($I_1 \neq 0$ and $I_1\neq K_2$), we have two trivial solutions: $\Delta\varpi= 0$ and $\Delta\varpi= \pi$. 
These solutions are often referred as Mode I ($\Delta\varpi=0$) and Mode II ($\Delta\varpi=\pi$). 
In mode I, the lines of apses of the two planets are aligned having the periapses on the same side; 
In mode II, the situation is similar but the two periapses are in opposite directions (the periapses are anti-aligned). Ordinary motions are oscillations around these fixed points.

\begin{figure}[t]
\centerline{
\epsfig{file=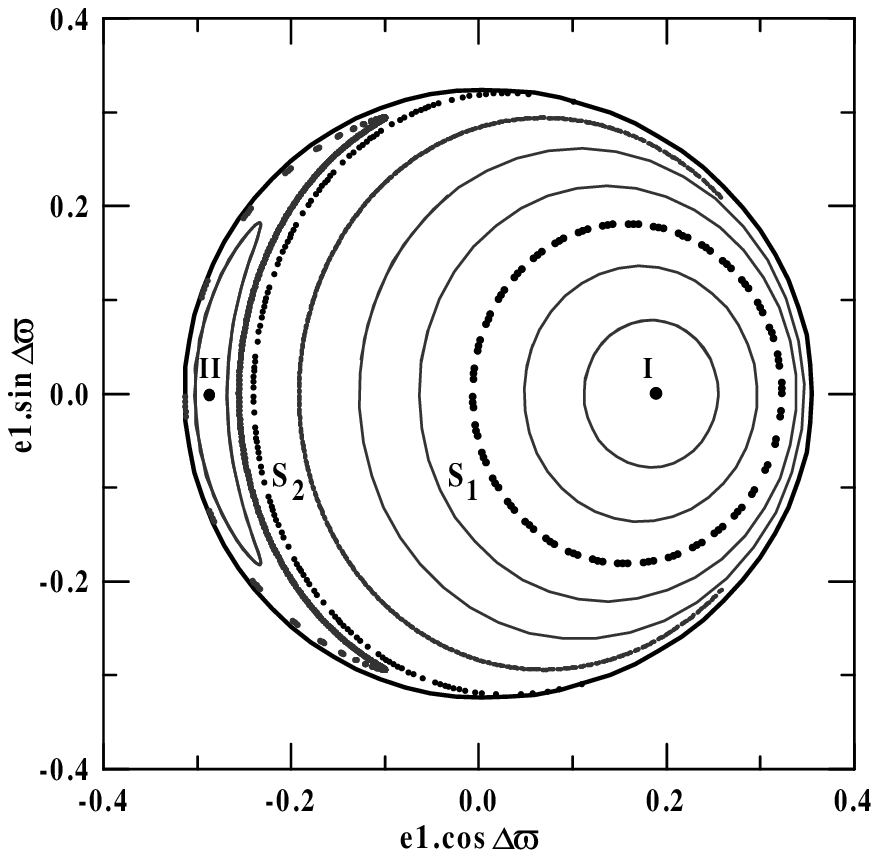,height=5.8cm,clip=}\hspace*{2mm}
\epsfig{file=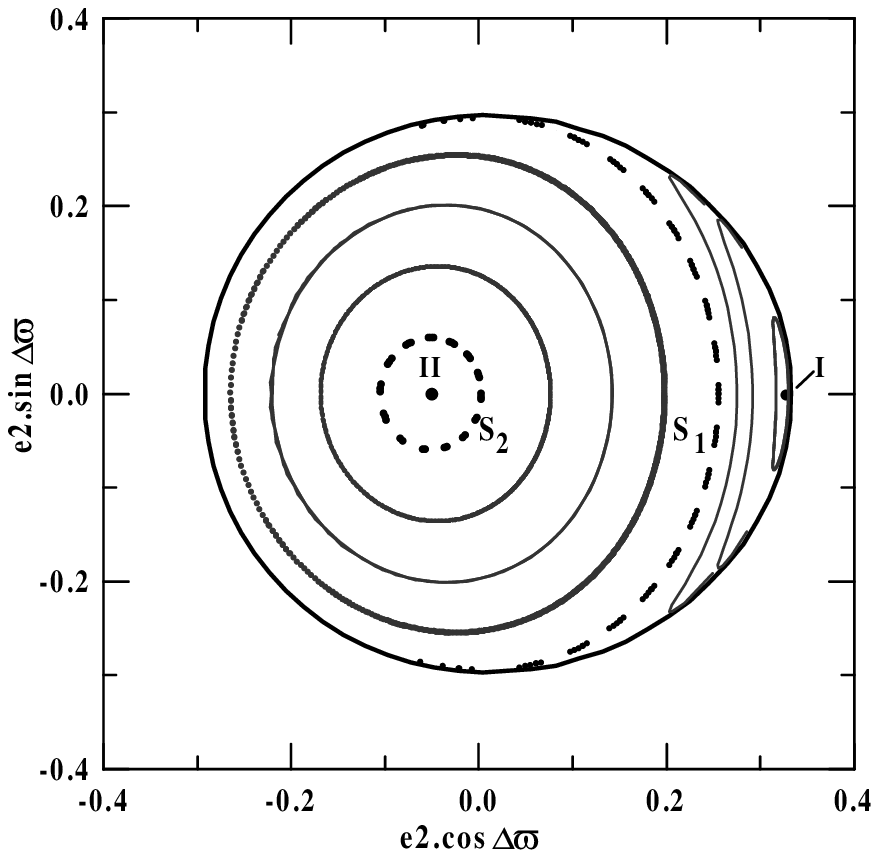,height=5.8cm,clip=}}
\caption{Secular variations of a system of planets with the same masses and semi-major axes as planets $c$ and $d$ of $\upsilon$And. The outer curves show the boundary of the energy manifold. The dashed curves $S_1$ and $S_2$ represent motions through the singularities of eqns.\,(\ref{onedeg}) (see text). The actual motion of planets $\upsilon$And $d$ and $\upsilon$And $c$ (see fig. \ref{ups}) is an oscillation around Mode I fixed point.}
\label {upsAnd}
\end{figure}

Solutions of the above equations corresponding to the masses, semi-major axes and energy level of the planets $c$ and $d$ of $\upsilon$And are shown in Fig.\,\ref{upsAnd}. They are presented in two different planes. One in which the coordinates are $e_1\cos\Delta\varpi,\, e_1\sin\Delta\varpi$ ($e_1$ is the eccentricity of $\upsilon$And $c$) and another, not independent, in which the coordinates are $e_2\cos\Delta\varpi,\, e_2\sin\Delta\varpi$
($e_2$ is the eccentricity of $\upsilon$And $d$).
 
On each figure, we can see the two fixed points above called Mode I and Mode II. In the left-hand phase plane, corresponding to the eccentricity of planet $c$, motions around the Mode I fixed point dominate; the Mode II fixed point lies near the left-hand boundary of the energy surface. In the right-hand figure, corresponding to planet $d$'s eccentricity, the situation is reversed and the flow is dominated by motions around the Mode II fixed point which lies near the center.
(For a discussion on the periodic orbits corresponding to the fixed points, see Michtchenko \& Ferraz-Mello, 2001; Michtchenko and Malhotra, 2004.)

It is important to note that even though the motion of angle $\Delta\varpi$ may be either an oscillation (about $0$ or $180^o$) or a circulation, there is no separatrix associated with an unstable infinite-period solution separating these motions. 
To better understand this feature, we plot by dashed lines two special solutions on each figure. 
These solutions are associated with the singularities in eqns.\,(\ref{onedeg}), which take place at $I_1=0$ and $I_1=K_2$ (that is, $I_2=0$).
One of these solutions, presented by the curve $S_1$, was calculated with initial condition $I_1 \simeq 0$ and is seen as a smooth curve passing through the origin on the left-hand side figure. At variance, $S_2$, calculated with initial condition $I_2\simeq 0$ is seen as the `false' separatrix between the domains of the motion around the two different fixed points. An analogous situation is seen in the right-hand-side figure where, now, $S_2$ is a smooth curve passing through the origin and $S_1$ separates the domains of the motion around the two different fixed points.
The motion along these separatrix-like curves is such that, when the representative point in one plane passes through the origin, in the other plane it is at the boundary of the separatrix-like curve and jumps from one boundary to another. 
However, such jump does not mean that the motion is passing through a singularity. 
It is just the result of the topological inadequacy of the plane to represent these solutions; they would be better drawn over a sphere (see Pauwels, 1983).

Figure \ref{upsAnd} shows some important features of the secular motion of two planets. In solutions close to Mode I (the right-hand side fixed point), the secular angle $\Delta\varpi$ oscillates about $0$ and the planet eccentricities undergo small oscillations about the value corresponding to Mode I equilibrium. In a similar way, the solutions close to Mode II (the left-hand side fixed point), the secular angle $\Delta\varpi$ oscillates about $180^o$. 
At mid-way from Mode I to Mode II, there is a large region of the phase space, corresponding to solutions where the motion of the secular angle $\Delta\varpi$ is a \underline{prograde} circulation. 

The motions around Mode I and Mode II are two opposite stable ways of the planetary system to be aligned.
In Mode II, the closest approaches between the planets occur when $\upsilon$And $c$ is at apoapsis and $\upsilon$And $d$ at periapsis, simultaneously. This situation can never occur in Mode I.

Fig. \ref{upsAnd} is akin to surfaces of section of the two-degrees-of-freedom system. The curves in each plane are defined by initial conditions on the plane plus one condition out of the plane (the other eccentricity, or, equivalently, $K_2$), which is adjusted in such a way that all curves correspond to the same energy. Therefore, it is not a phase portrait. (Phase portraits of one-degree-of-freedom Hamiltonian are sets of trajectories of different energies. See the phase portraits of $<H>$ in Pauwels, 1983.) 
This choice makes fig. \ref{upsAnd} more suitable for comparison to similar plots obtained for 2-planet resonant systems (Michtchenko \& Ferraz-Mello, 2001; Callegari Jr. et al., 2004).  

\section{Resonant Dynamics}
In the previous section, the Hamiltonian was averaged over the two mean longitudes $\lambda_1$ and $\lambda_2$.
This procedure is not valid if the two planets have commensurable periods, since, in this case, $\lambda_1$ and $\lambda_2$ are no longer independent:
\bd
\frac{p+q}{p} \ {\rm resonance} \Longleftrightarrow\frac{T_2}{T_1}\simeq\frac{p+q}{p}
\ed
The averaging over the two longitudes will kill all terms depending on the longitudes including those depending on the critical combination
\bd
(p+q)\lambda_2 - p\lambda_1.
\ed
However, these terms play a major role in the dynamics of the two planets and should remain in $<H>$.
To preserve them, we define, before the averaging, the following set of planar canonical variables:

\be\begin{array}{rcl}
\lambda_1  &\hspace*{0.5cm}&  J_1 = L_1 + s (I_1+I_2)\\
\lambda_2  & & J_2 = L_2 - (1+s) (I_1+I_2)\\
(1+s)\lambda_2 - s\lambda_1 - \varpi_1 = \sigma_1 & 
& I_1 = L_1 - G_1  \\
(1+s)\lambda_2 - s\lambda_1 - \varpi_2 = \sigma_2 && 
 I_2 = L_2 - G_2
\label{MNeq29}\end{array}\ee
where $s=p/q$. The two angular variables $\sigma_i$ are the critical angles. With the angles thus introduced, the generic argument $m\ell_1 - n\ell_2 + l \Delta \varpi$ of the disturbing function becomes
$(m-l)\sigma_1 - (n-l)\sigma_2 + [m(1+s)-ns] (\lambda_1-\lambda_2)$. Note that, because of the invariance of $H_1$ to rotations, the mean longitude only appears through the mean synodic longitude $\lambda_1-\lambda_2$. It is easy to see that the ``action'' conjugate to the missing angle is the total angular momentum ${\cal L} = G_1+G_2 = (L_1-I_1)+(L_2-I_2) = J_1 + J_2$. The averaging over the mean longitudes (or over the mean synodic longitude) can, now, be done and the critical angles will be preserved inside $\sigma_1$ and $\sigma_2$.
	
After the averaging,
\bd
<H>\; = - \sum_{i=1}^2 \frac {\mu_i^2 \beta_i^3}{2L_i^2} - 
R_{\rm res}
\ed
where
\bd
R_{\rm res} = \frac{Gm_1m_2}{a_2} \sum_{i,j,k,m',n'} C_{[\cdots]} \alpha^i e_1^j  e_2^k \cos [m' q \sigma_1 + n' (\sigma_2-\sigma_1)]
\label{Rres}
\ed

The momenta whose conjugate angles no longer appear in $<H>$ are first integrals (only 2, now):
\bea
J_1&=&{\rm const.}\nonumber \\
J_2&=& {\rm const.}
\eea
$J_1+J_2=G_1+G_2$ is the Angular Momentum, whose conservation in the system before the averaging was discussed. 
It is worth emphasizing the fact that the $L_i$ (i.e. the semi-major axes) are no longer invariant.

The two integrals above may be combined to give
\be
(1+s)L_1+sL_2 = {\rm const.}
\ee
(Sessin and Ferraz-Mello, 1984). 
This integral of the resonant dynamics means that $a_1$ and $a_2$ vary in anti-phase. When one of the semi-axis increases, the other necessarily decreases. 

The above variables may also be combined to give:
\be
AMD = I_1+I_2 = {\rm const.} + \frac{L_1}{s}.
\ee
The AMD also is no longer invariant, but its variation is small and thus limitations of the eccentricities similar to that observed in the secular motion (but different) exist.

\subsection{Resonant Stationary Solutions. Apsidal Corotation.}

The averaged system is, now, an irreducible two-degrees-of-freedom system. An important feature of this system is the existence of stationary solutions (Beaug\'e et al., 2003; Ferraz-Mello et al. 2003; Lee and Peale, 2003). These solutions are defined by the equations
\be
\frac{d I_i}{dt}=\frac{\partial <H>}{\partial \sigma_i}  =  
\frac{\partial R_{\rm res}}{\partial \sigma_i } =0,
\hspace{2cm}
\frac{d\sigma_i}{dt}=\frac{\partial <H>}{\partial I_i}  =  0.
\label{stationary}\ee
They are such that $I_i$ and $\sigma_i$ are constant (except for the short period terms eliminated by the averaging and for contributions of higher orders). 
Constant $I_i$'s mean semi-major axes and eccentricities constant in these solutions; 
$\sigma_1$ and $\sigma_2$ constant mean that $\Delta\varpi=\sigma_1-\sigma_2$ is constant, that is, the periapses are moving with same velocities so that their mutual separation do not vary. 
This frozen relative state in resonant systems is known as {\it apsidal corotation}.

Equations (\ref{stationary}) may be studied separately.
The first equation says that the stationary solutions lie at the extrema of the function $R_{\rm res}$ with respect to the variables $\sigma_i$. 
These extrema depend only on the ratio of the masses of the two planets and on the eccentricities (constants in the stationary solution).
The factor $\frac{Gm_1m_2}{a_2}$ does not affect the results. 

\begin{figure}[h]
\centerline{
\epsfig{file=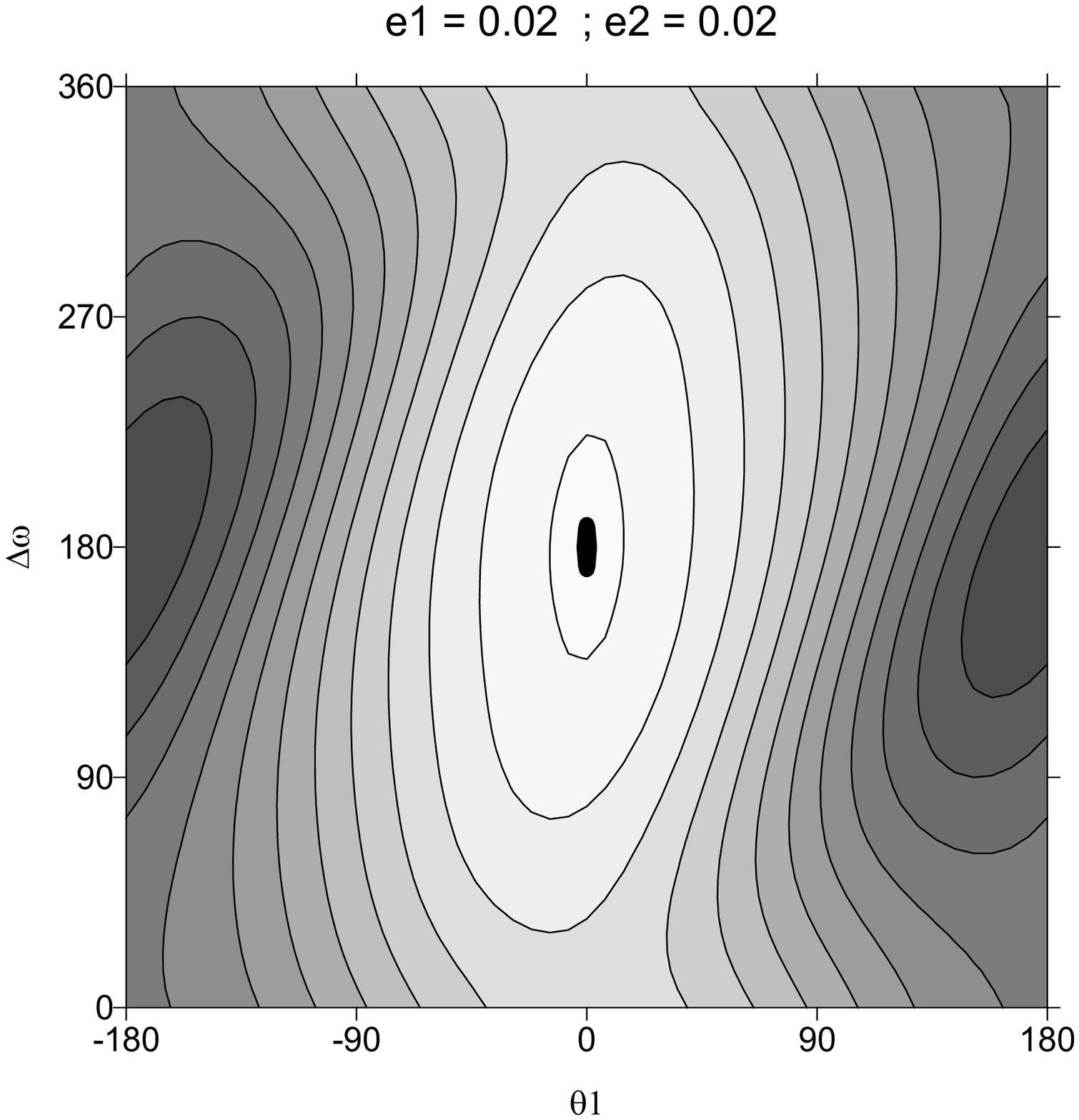,height=6.3cm,clip=}
\epsfig{file=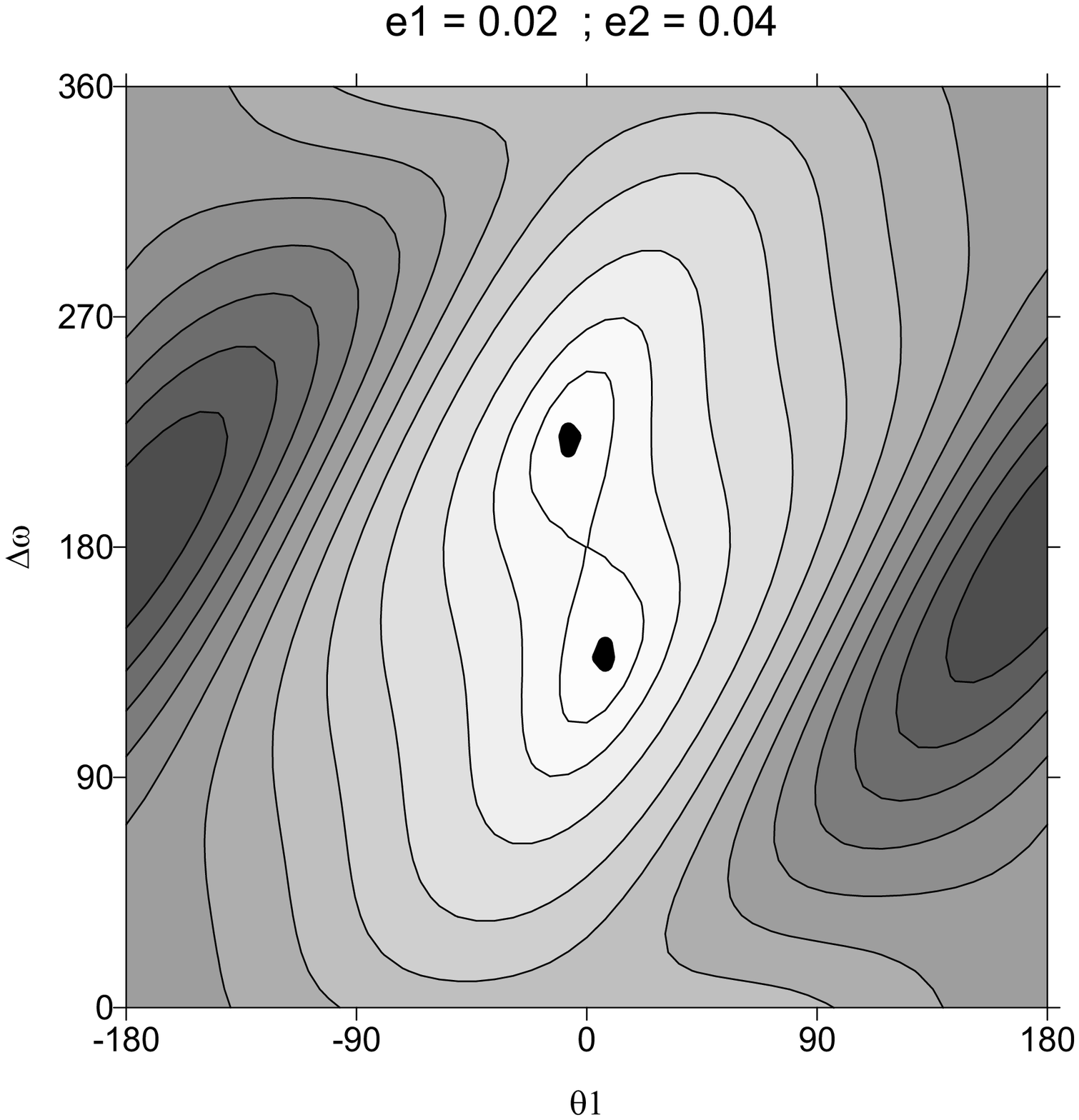,height=6.3cm,clip=}}
\centerline{
\epsfig{file=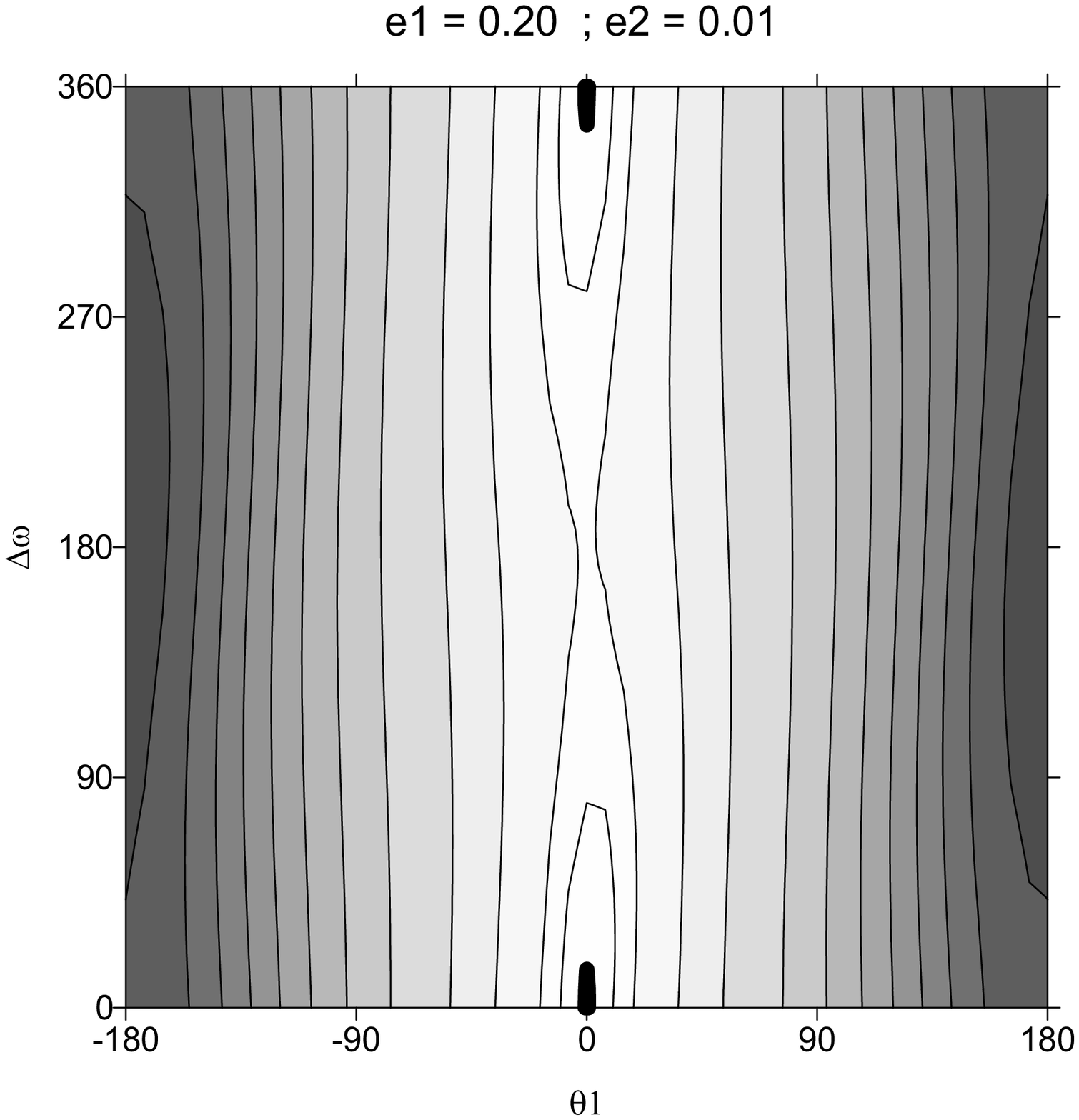,height=6.3cm,clip=}
\epsfig{file=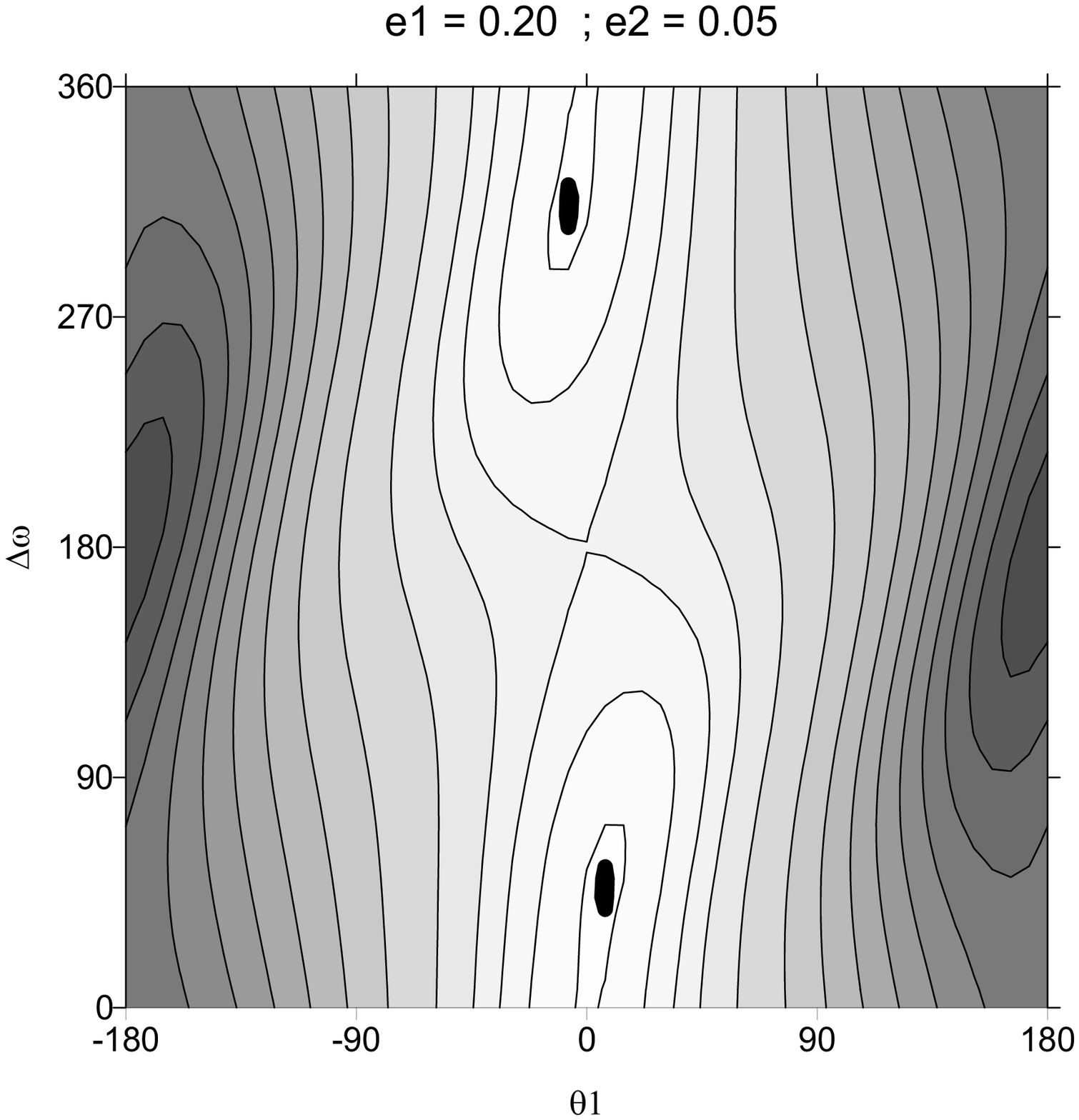,height=6.3cm,clip=}}
\caption{Contour plots of $R_{\rm res}$ in the 2/1 resonance for 4 given pairs of eccentricity values. Abcissas: $\sigma_1=2\lambda_2-\lambda_1-\varpi_1$; Ordinates: $\Delta\varpi=\varpi_1-\varpi_2$.}
\label{hcte}
\end{figure}

Figures \ref{hcte} are contour plots of the function $R_{\rm res}$ for given  $e_1,\,e_2$ and $\alpha$ (taken at $\alpha=a_1/a_2\simeq 0.63$, value corresponding to the resonance 2/1).
For the sake of an easier interpretation, we used the angles $\sigma_1,\Delta\varpi=\sigma_2-\sigma_1$, instead of $\sigma_1, \sigma_2$. 
The extrema seen in these figures may correspond to stable stationary solutions or not. $<H>$ is a function of 4 variables and only 2 variables are considered in these figures. Therefore, what appears as an extremum in this picture is not necessarily one extremum in the full phase space. The stable solutions considered in this paper are those corresponding to the centers in the white areas. However, one should be aware that they are not the only stable stationary solutions in this problem (see Hadjidemetriou and Psychoyos, 2003).

The two first plots correspond to low $e_1\,$ ($e_1=0.02$).
For small $e_2\,$ ($e_2=0.02$) the extremum corresponding to stable solutions is such that $\Delta\varpi=\pi$ ($\sigma_1=0,\sigma_2=\pi$).
In this solution, the periapses are anti-aligned.
When $e_2$ is larger ($e_2=0.04$ in the right-hand plot), the extremum seen in the left-hand plot becomes a saddle point and a bifurcation gives rise to two extrema symmetric with respect to the saddle. 
These extrema correspond to asymmetric stationary solutions where $\Delta\varpi=\sigma_2-\sigma_1$ remains constant but with a value not necessarily equal to zero or $\pi$ or commensurable with $\pi$.
The second row of plots correspond to high $e_1\,$ ($e_1=0.2$).
For small $e_2\,$ ($e_2=0.01$) the extremum corresponding to stable solutions is such that $\Delta\varpi=0$ ($\sigma_1=\sigma_2=0$).
In this solution, the periapses are aligned.
When $e_2$ is larger ($e_2=0.05$ in the right-hand plot), the same phenomenon seen in the first row occurs: the extremum seen for low $e_2$ gives rise to two extrema symmetric with respect to the saddle. 
As in the previous case, these extrema correspond to asymmetric stationary solutions.
These asymmetric solutions, depending on the eccentricities, may be found on a large set of points $\sigma_1,\Delta\varpi$. 
\begin{figure}[h]
\centerline{
\epsfig{file=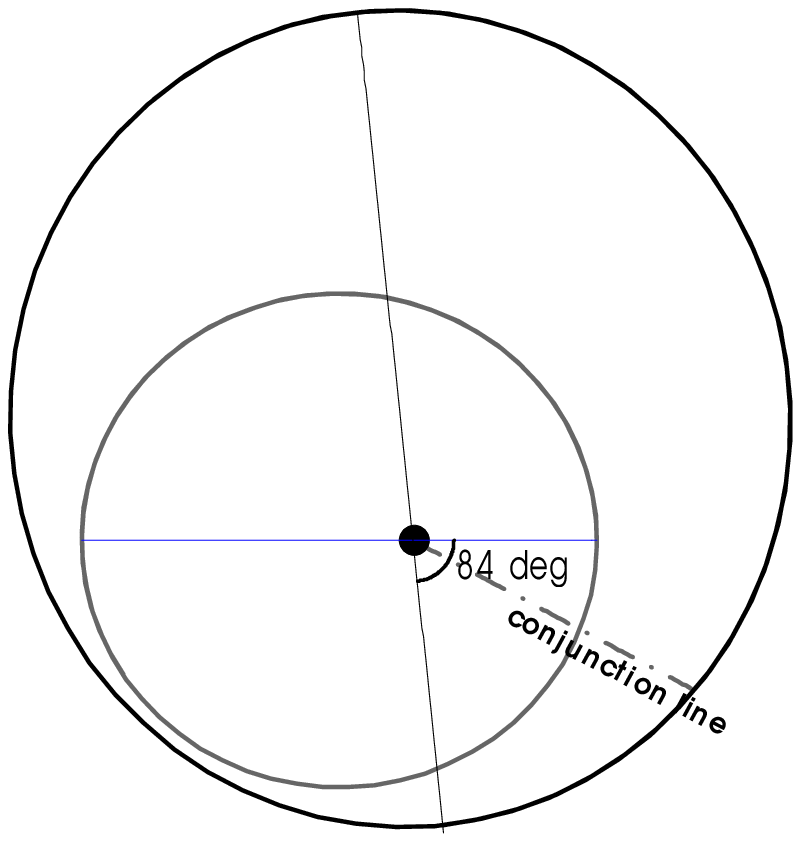,height=6cm,clip=}\hspace{1cm}
\epsfig{file=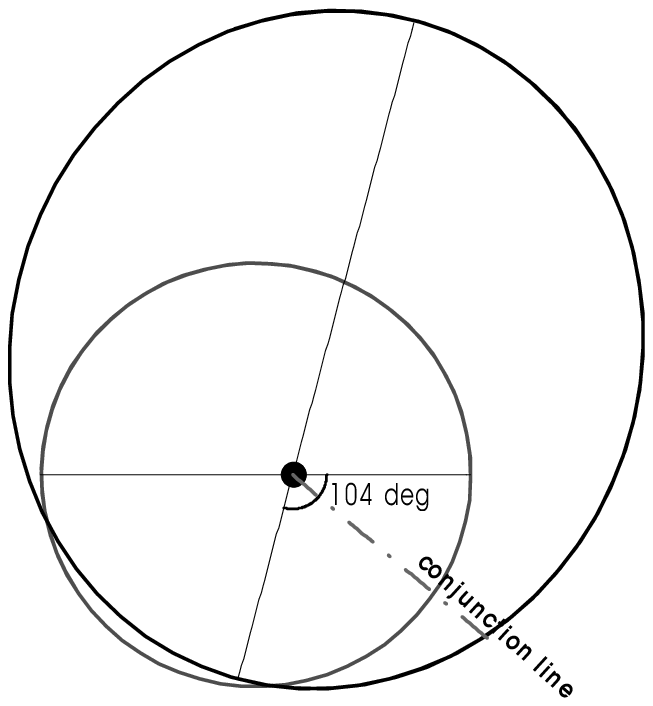,height=6cm,clip=}}
\caption{Asymmetric stationary solutions. 
{\it Left:} $|\Delta\varpi| = 84^\circ$, $\, e_1=0.286$ and $e_2=0.3$. 
{\it Right:} $|\Delta\varpi| = 104^\circ$, $\,e_1=0.17$ and $e_2=0.38$.}
\label{Nanj13}
\end{figure}

To complete the determination of the stationary solution, we need to solve the remaining equation:
\be
\frac{\partial <H>}{\partial I_i} = s n_1 - (1+s) n_2 + \frac{\partial R_{\rm res}}{\partial I_i} = 0.
\label{segunda}\ee
At variance with the previous equation, the solutions of this equation depend on the masses. However, it is easy to see that it depends almost only on the ratio of the masses of the two planets. Indeed, the $n_i$ are constants in the stationary solutions and the commensurability relation at the resonance $\frac{p+q}{p}$ is 
\bd
s n_1 - (1+s)n_2 = 0; 
\ed
In the remaining part, the derivatives of $R_{\rm res}$ 
with respect to $I_i$ change the dependence on the masses. We remember that
\bd
I_i=m_i (1+\frac{m_i}{m_0})^{-1/2} \sqrt{Gm_0a_i} (1-\sqrt{1-e_i^2}).
\ed
Thus, the coefficient in front of the summation in $R_{\rm res}$ becomes linear in the planet masses after the derivative with respect to $I_i$.
Therefore, eqn.\,(\ref{segunda}) has the form $A_1m_1+A_2m_2 = 0$, whose solutions do not depend on the masses themselves but only on the mass ratio $m_2/m_1$. This is not a rigorous statement. In fact, the semi-major axes and eccentricities are functions of $I_i$ that include also the factor $m_0+m_i$. This means that $A_1$ and $A_2$ are independent on the masses only in a first approximation. Even if their variation with the masses is small for the range of masses of the considered planets, this variation exists and will affect the solutions in case of large ratios $m_i/m_0$. Beaug\'e {\it et al.} (2003) have shown that the stationary orbits obtained in this section exist for planet masses less than $\sim 10^{-2}$ of the star mass. 

\begin{figure}[h]
\centerline{
\epsfig{file=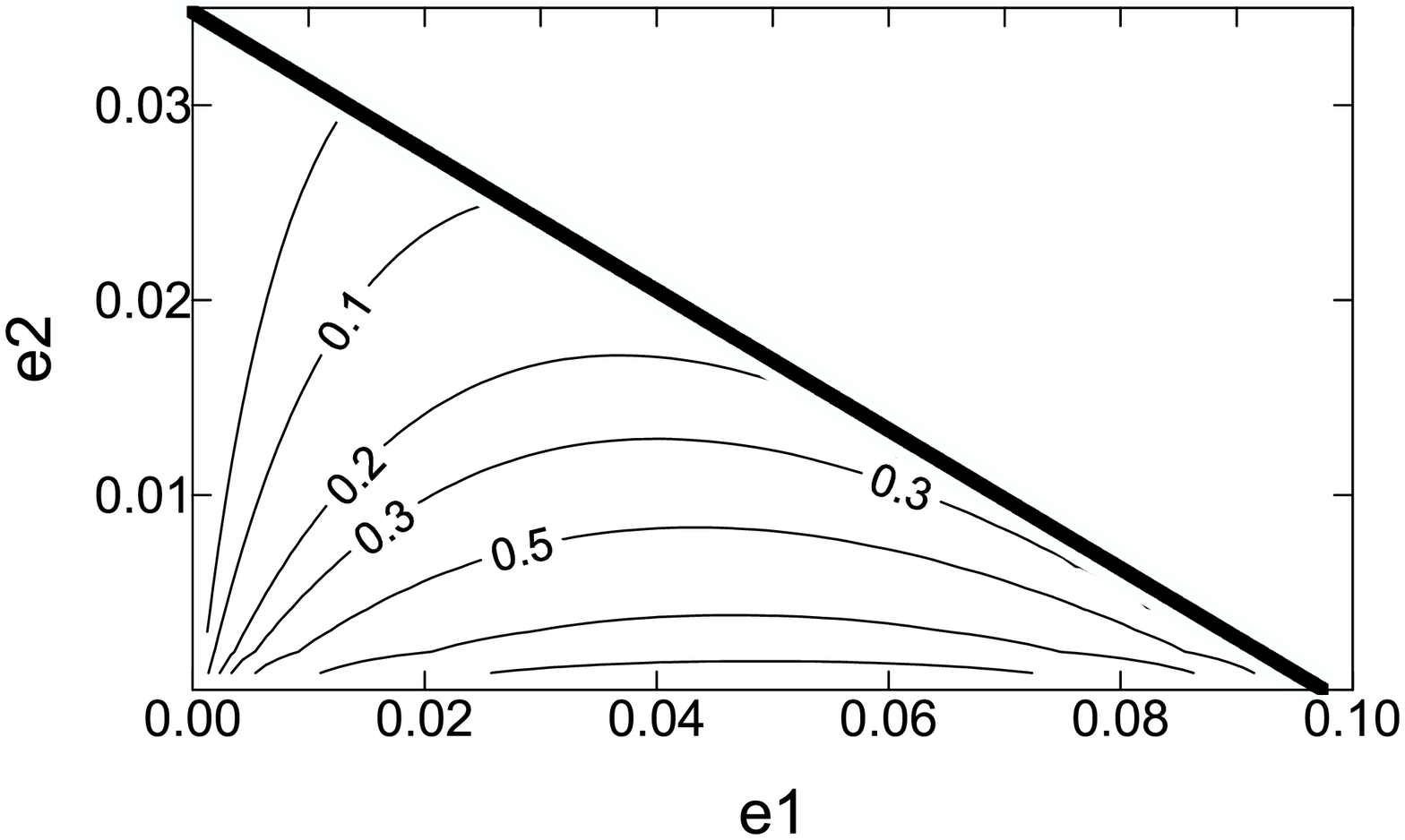,height=3.8cm,clip=}
\epsfig{file=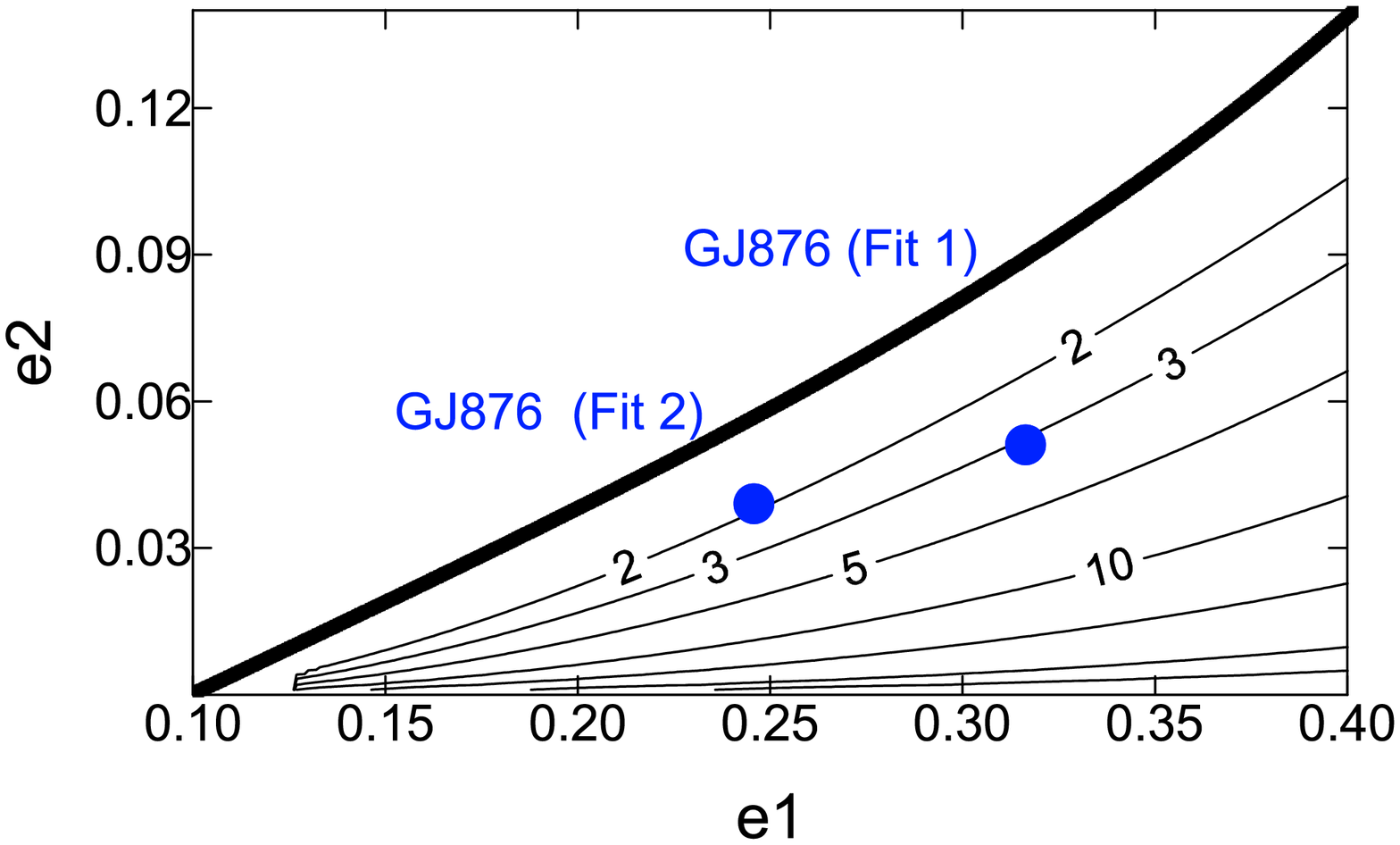,height=3.8cm,clip=}}\vspace*{3mm}
\centerline{
\epsfig{file=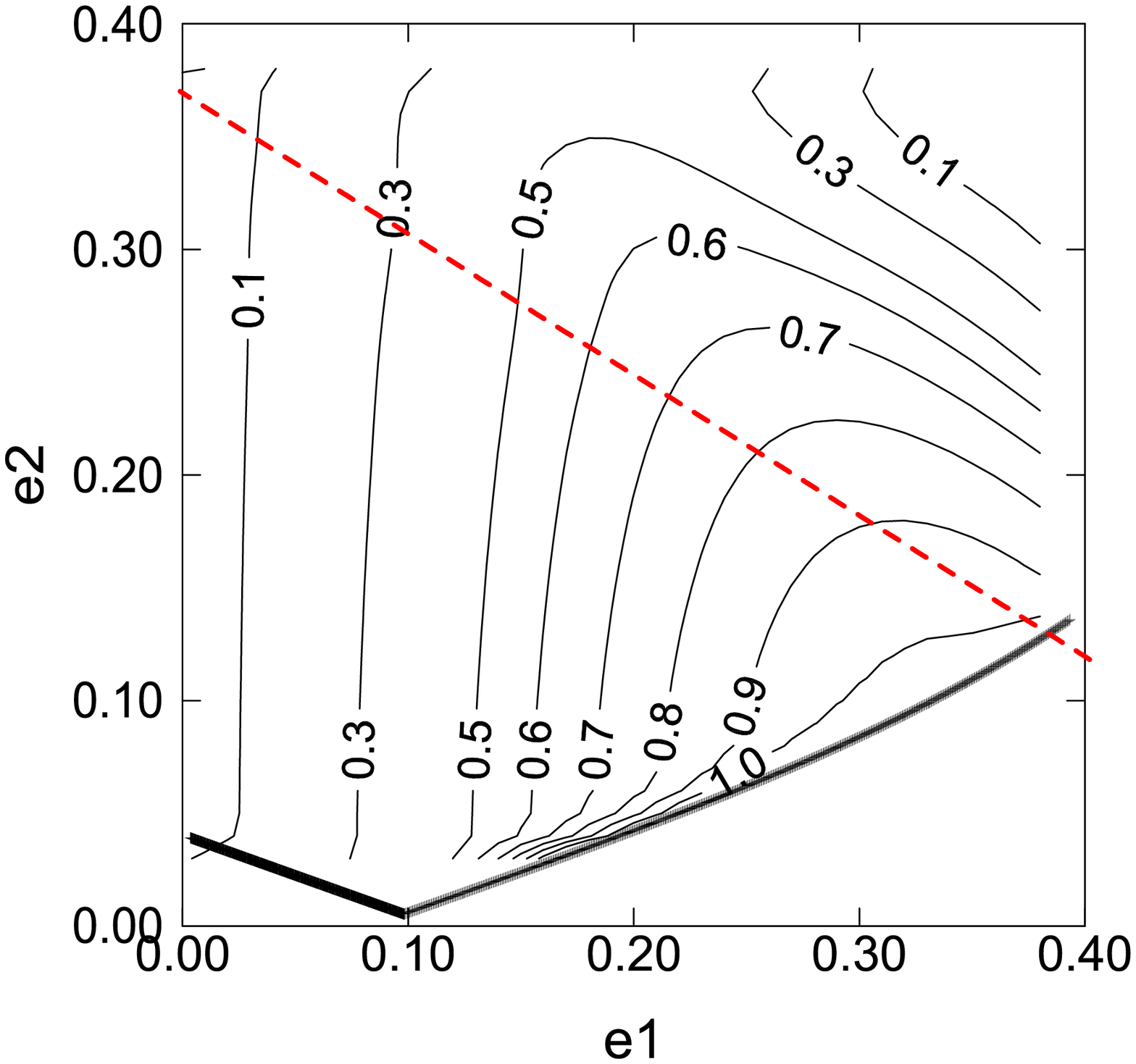,height=6cm,clip=}}
\caption{Loci of the stationary corotation solutions of the 2/1 resonance for several mass ratios $m_2/m_1$. Top figures correspond to the symmetric solutions of the two left-hand side plots in fig. \ref{hcte}. The points corresponding to two early determinations of the elements of Gliese 876 are shown in one of these plots. The bottom figure corresponds to the asymmetric solutions of the two right-hand side plots in fig. \ref{hcte}. The line across these curves shows the values of the eccentricities for which $0.63(1+e_1)=(1-e_2)$. In all panels, the thick line shows the boundary between the domains of symmetric and asymmetric solutions.} 
\label{Apjfigs}
\end{figure}

The above equations were used to find apsidal corotation solutions in the case of planets in 2/1 and 3/1 mean-motion resonances. The relationship between eccentricities and mass ratios in some of these solutions are shown in fig. \ref{Apjfigs}. The top panels correspond to symmetric solutions. In the left-hand side panel, the periapses are anti-aligned. This is the case of the two innermost Galilean satellites of Jupiter: Io and Europa. In the right-hand side panel, the periapses are aligned. This is the case of the two planets in orbit around the star Gliese 876. The thick lines in the two top panels show the boundary above which symmetric solutions no longer exist. At the thick line, the solutions bifurcate into pairs of asymmetric solutions. The relationship between eccentricities and mass ratios in the domain of asymmetric solutions is shown in the bottom panel. It is worth noting that the mass ratio $m_1/m_2$ in the bottom panel is always smaller than a limit close to 1.0. This situation is often called ``exterior case'' since it corresponds to have the smaller body in an orbit exterior to that of the more massive one. Asymmetric apsidal corotations are known in the exterior asteroidal case (Beaug\'e, 1994). Asymmetric periodic solutions in the restricted three-body problem were first shown to exist by Message (1958). We may also mention a similar behavior, in deep resonance, of the Laplacian critical angle of the Galilean satellites of Jupiter: $\lambda_1-3\lambda_2+2\lambda_3$ (Greenberg, 1987).

\section{Capture into Resonance}\label{capture}

In this section, we present the results of a series of numerical simulations of the dynamical evolution of fictitious pairs of planets under the action of a non-conservative perturbation that adds angular momentum and energy to the orbit of the innermost planet. The planets are small (some $10^{-5}$ of the central body mass) and the mass ratio is $m_2/m_1=0.538$ (i.e., the so-called exterior case). The actual calculations were done with satellites instead of planets, but the physical nature of the system does not matter in the following discussion. The physics and used methodology are in (Ferraz-Mello et al., 2003). 

The initial distances to the star are just behind the 2/1 resonance: $\alpha=a_1/a_2 =0.612$. When the semi-major axis of $m_1$ increases, $a_1$ increases and the mean-motion resonance ($\alpha=0.63$) between $m_1$ and $m_2$ is reached. Capture then can take place. The probability of capture depends on the rate of variation of $a_1$ -- if the rate is high, the orbit crosses the resonance without capture, one phenomenon very well studied in the case of one massless particle. Other factors influencing the probability of capture are the orbital eccentricities -- capture is more probable when orbital eccentricities are small (Dermott et al., 1988; Gomes, 1995). In our calculations, initial eccentricities were lower than 0.001 and the physical parameters were adjusted to have slow resonance approximation. 
Figure \ref{tide5-a} shows the evolution of the semi-major axes. 

\subsection{Capture into Apsidal Corotation}\label{apsidal}
\begin{figure}[h]
\figeps{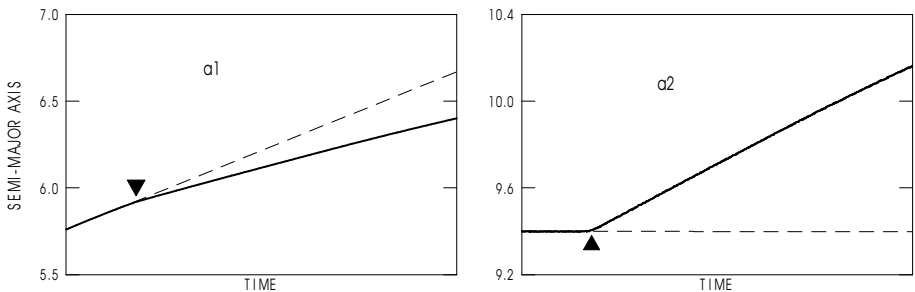}{3.8cm}
\caption{Evolution of the semi-major axes before and after the capture into resonance. Triangles mark the moment of the capture. Dashed lines extrapolate the evolution before the capture and show the change in slope of the evolution lines. (arbitrary units)}
\label{tide5-a}
\end{figure}

The system evolves with the innermost orbit receding from the central body (because of the non-conservative forces acting on $m_1$) up to the moment where the system is captured into a resonance. 
$a_2$ is almost constant. 
When the 2/1-resonance is reached, the system is trapped by the resonance. 
As known since Laplace, after the capture, $m_1$ continuously transfers one fraction of the energy that it is getting from the non-conservative source to $m_2$, so that $a_2$ also increases. One may note from fig. \ref{tide5-a} that, after the capture into the resonance, $a_1$ increases at a smaller pace than before the capture. The increase of the semi-major axes is such that the ratio $a_1/a_2$ remains constant.  

\begin{figure}[h]
\centerline{
\epsfig{file=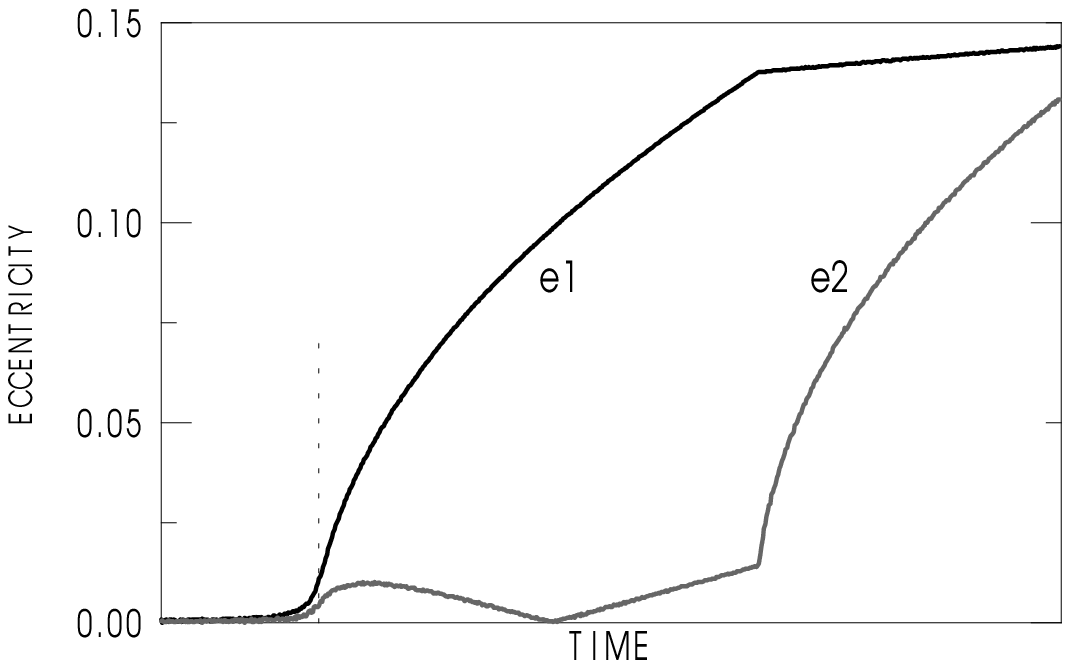,height=6cm,clip=}}
\centerline{
\epsfig{file=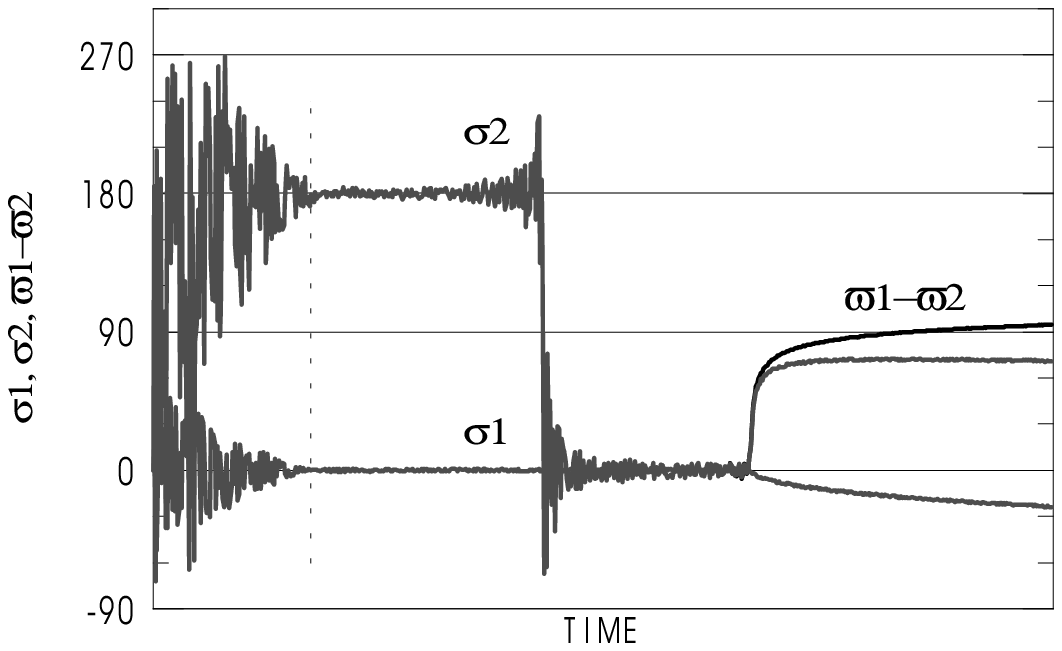,height=5.85cm,clip=}}
\caption{Variation of the eccentricities, critical angles ($\sigma_i$) and $\Delta\varpi$ in the same time interval as figure \ref{tide5-a}. 
The vertical dotted lines show the moment of the capture. $\Delta\varpi$ is only shown in the final part since it does not differ significantly from $\sigma_2$ in the time interval between the capture into resonance and the bifurcation.} 
\label{tide5-esig}
\end{figure}

Figures \ref{tide5-esig} show the variation of the eccentricities, critical angles $\sigma_i=2\lambda_2-\lambda_1-\varpi_i$  and $\Delta\varpi$ in the same time interval as the previous figures. 
They show that, after capture, the two critical angles become trapped in the neighborhood of 0 and $\pi$, respectively and, consequently, the angle $\Delta\varpi$ is trapped in the neighborhood of $\pi$. The capture into a symmetric apsidal corotation with anti-aligned periapses is thus simultaneous with the capture into the resonance.

\subsection{Evolution after Capture}
Figure \ref{tide5-esig} also shows that, after some time, $\sigma_2$ jumps from $\pi$ to 0 and the apsidal corotation becomes one with aligned periapses. This change is not the result of a discontinuous process. The left-hand side plot shows that the change happens when the eccentricity $e_2$ is zero. Thus, we may describe the process by a momentary circularization of the orbit such that, when it becomes an ellipse again, the periapses is not at the same side as before. The large transients shown by the variation of the angle $\sigma_2$ are just due to the sensibility of the angle $\varpi_2$ to small changes when $e_2\sim 0$. 

The apsidal corotation with aligned periapses does not last long. The figures show that the angles depart from zero and the apsidal corotation becomes asymmetric. At this moment, there is a discontinuity in the rates of variation of the eccentricities (the elbows seen in the curves $e_i(t))$.

Figure \ref{tide5-edpi} shows the evolution over a time interval almost 10 times longer. The first point to stress here is that such a time span is likely beyond physical signification. Each panel of fig. \ref{tide5-edpi} combines the variations of $e_1$ (resp. $e_2$) and $\Delta\varpi$ in a same plot in polar coordinates in which the radius vector is the eccentricity and the polar angle is $\Delta\varpi$.
We itemize the important points to be noted:

\begin{figure}[h]
\centerline{
\epsfig{file=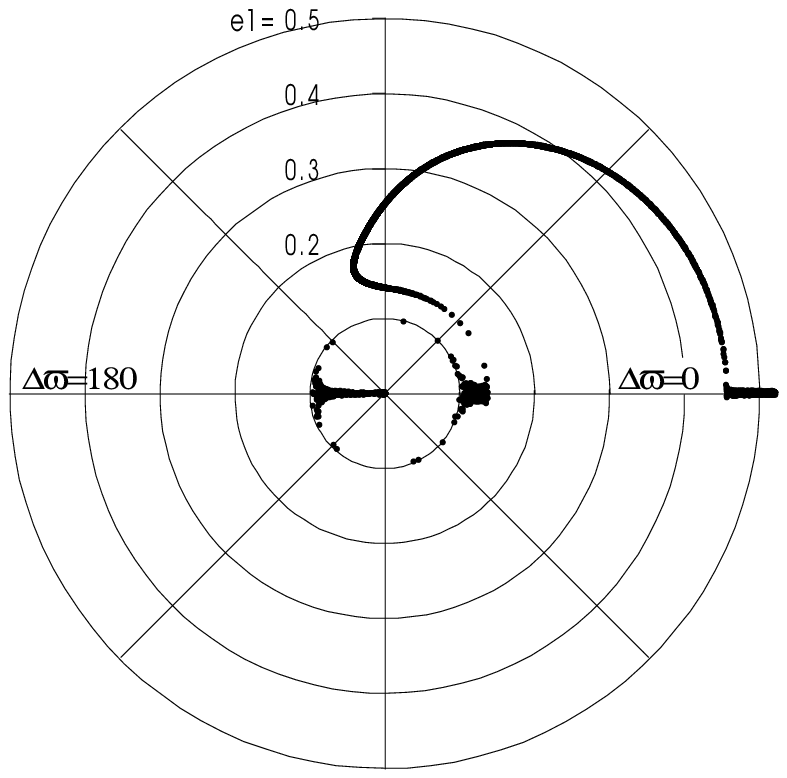,height=5.8cm,clip=}\hspace{.2cm}
\epsfig{file=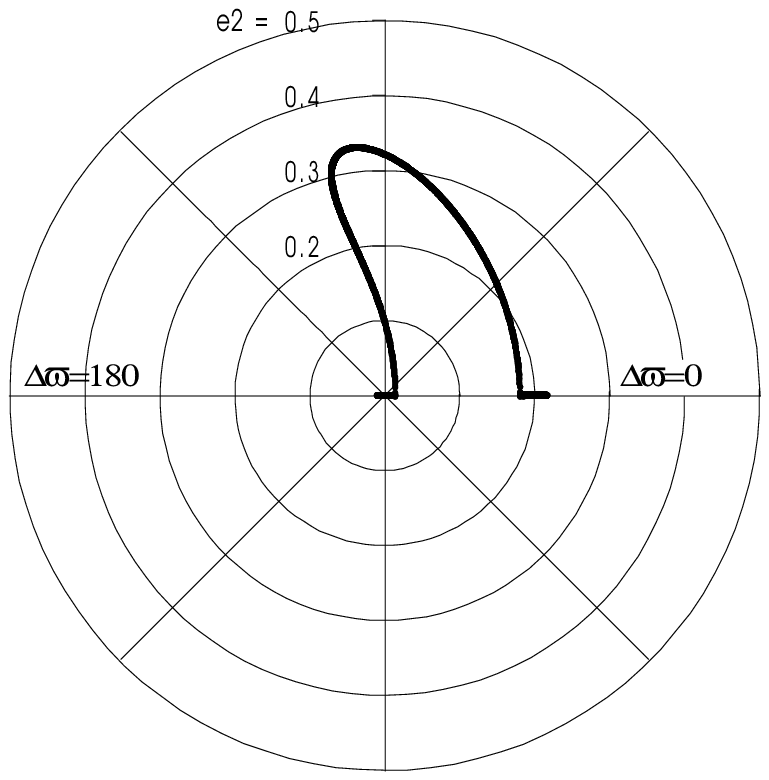,height=5.8cm,clip=}}
\caption{Polar plots in the planes $e_i\exp i\Delta\varpi$.} 
\label{tide5-edpi}
\end{figure}

\begin{itemize}
\item[$\bullet$]
The eccentricity $e_1$ increases monotonically. When it reaches $\sim 0.46$, the asymmetric apsidal corotation changes back to a symmetric configuration with aligned periapses. 
\item[$\bullet$]
After the bifurcation, the followed solution lies in the upper half-plane. This is not the only possibility. The motion could have followed a mirror path in the lower half-plane. The probability of following one or another branch is the same. 
\item[$\bullet$]
The phenomenon leading to the transformation from anti-aligned to aligned periapses is clearly seen in the right-hand side panel, where the trajectory is seen crossing the origin of the plot.
\end{itemize}

It is interesting to note that this picture has a counterpart in the study of periodic orbits of the 3-body problem. The study of symmetric periodic solutions shows the existence of two separated stable branches with aligned periapses; these two branches are tied with continuity by a branch of unstable periodic orbits (Hadjidemetriou, 2002). The unstable branch corresponds to the saddles shown in the fourth panel of fig. \ref{hcte}. In fig. \ref{tide5-edpi} it would appear as a right shortcut on the horizontal axis tying the initial and final segments of stable solutions with aligned periapses. 

\section{Conclusion}
The contents of this paper include with variable emphasis, the topics of a series of lectures whose main title was ``Routes to Order: Capture into Resonance''. This was indeed the subject of the last section above. The study of this subject has however shown that differently from the restricted three-body problem, the capture into resonance drives the system immediately to stationary solutions known as ``apsidal corotations''. The whole theory of these solutions was also included in the paper from the beginning, that is, from the formulation of the Hamiltonian equations of the planetary motions and the expansion of the disturbing function in the high-eccentricity planetary three-body problem. The secular theory of non-resonant systems was also given. Motions with aligned or anti-aligned periapses, resonant or not, resulting from non-conservative processes (tidal interactions with the disc) in the early phases of the system life, seem to be frequent in extra-solar planetary systems.

\end{document}